\RequirePackage{fix-cm}

\documentclass[smallextended]{svjour3}

\usepackage{natbib}
\usepackage{balance}
\usepackage{fancyhdr}
\usepackage{amssymb}
\usepackage{amsmath}
\usepackage[figuresright]{rotating}
\usepackage{wrapfig}
\usepackage{multirow}
\usepackage{graphicx}
\usepackage{algorithm}
\usepackage{algorithmic}
\usepackage{url}
\usepackage{booktabs}
\usepackage{color}
\usepackage{moreverb}
\usepackage{balance}
\usepackage{amssymb,amsmath}
\usepackage{wrapfig}
\usepackage{multirow}
\usepackage{graphicx}
\usepackage{algorithm}
\usepackage{algorithmic}
\usepackage{epstopdf}
\usepackage{tabularx}
\usepackage{bigstrut}
\usepackage{url}
\usepackage{enumitem}
\usepackage{miniplot}
\usepackage{ifthen}

\usepackage{pifont}
\usepackage{bbding}
\usepackage{booktabs}
\usepackage{threeparttable}
\usepackage{array}
\usepackage{ragged2e}
\usepackage{etoolbox}

\makeatletter

\patchcmd{\NAT@citex}
  {\@citea\NAT@hyper@{%
     \NAT@nmfmt{\NAT@nm}%
     \hyper@natlinkbreak{\NAT@aysep\NAT@spacechar}{\@citeb\@extra@b@citeb}%
     \NAT@date}}
  {\@citea\NAT@nmfmt{\NAT@nm}%
   \NAT@aysep\NAT@spacechar\NAT@hyper@{\NAT@date}}{}{}

\patchcmd{\NAT@citex}
  {\@citea\NAT@hyper@{%
     \NAT@nmfmt{\NAT@nm}%
     \hyper@natlinkbreak{\NAT@spacechar\NAT@@open\if*#1*\else#1\NAT@spacechar\fi}%
       {\@citeb\@extra@b@citeb}%
     \NAT@date}}
  {\@citea\NAT@nmfmt{\NAT@nm}%
   \NAT@spacechar\NAT@@open\if*#1*\else#1\NAT@spacechar\fi\NAT@hyper@{\NAT@date}}
  {}{}

\makeatother

\newboolean{showcomments}
\setboolean{showcomments}{true}
\ifthenelse{\boolean{showcomments}}
{ \newcommand{\mynote}[2]{
    \fbox{\bfseries\sffamily\scriptsize#1}
    {\small$\blacktriangleright$\textsf{\emph{#2}}$\blacktriangleleft$}}}
{ \newcommand{\mynote}[2]{}}

\ifthenelse{\boolean{showcomments}}
{
  
}{
  \newcommand{\mynote}[2]{}
}

\begin{document}

\title{Maintenance-Related Concerns for Post-deployed Ethereum Smart Contract Development: Issues, Techniques, and Future Challenges}


  \author{Jiachi Chen   \and Xin Xia \thanks{Xin Xia is the corresponding author.}
  	\and David Lo \and
	John Grundy \and Xiaohu Yang}


\institute{Jiachi Chen \at
	Faculty of Information Technology, Monash University, Australia\\
	\email{jiachi.chen@Monash.edu}
	\and
	Xin Xia \at
	Faculty of Information Technology, Monash University, Australia \\
	\email{xin.xia@monash.edu}
	\and
	David Lo \at
	School of Information Systems, Singapore Management University, Singapore\\
	\email{davidlo@smu.edu.sg}
	\and
	John Grundy \at
	Faculty of Information Technology, Monash University, Australia \\
	\email{John.Grundy@monash.edu}
	\and
	Xiaohu Yang \at
	College of Computer Science and Technology, Zhejiang University, China\\
	\email{yangxh@zju.edu.cn}}

\date{Received: date / Accepted: date}
\maketitle
\maketitle
\thispagestyle{fancy}
\lhead{}
\chead{}
\rhead{}
\lfoot{}
\cfoot{}
\cfoot{\thepage}
\renewcommand{\headrulewidth}{0pt}
\renewcommand{\footrulewidth}{0pt}
\pagestyle{fancy}
\cfoot{\thepage}

  \begin{abstract}
			
			Software development is a very broad activity that captures the entire life cycle of a software, which includes designing, programming, maintenance and so on. In this study, we focus on the maintenance-related concerns of the post-deployment of smart contracts. Smart contracts are self-executed programs that run on a blockchain. They cannot be modified once deployed and hence they bring unique maintenance challenges compared to conventional software. According to the definition of ISO/IEC 14764, there are four kinds of software maintenance, i.e.,  corrective, adaptive, perfective, and preventive maintenance. This study aims to answer (i) What kinds of issues will smart contract developers encounter for corrective, adaptive, perfective, and preventive maintenance after they are deployed to the Ethereum? (ii) What are the current maintenance-related methods used for smart contracts? To obtain the answers to these research questions, we first conducted a systematic literature review to analyze 131 smart contract related research papers published  from 2014 to 2020. Since the Ethereum ecosystem is fast-growing, some results from previous publications might be out-of-date and there may be a gap between academia and industry. To address this, we performed an online survey of smart contract developers on Github to validate our findings and received 165 useful responses. Based on the survey feedback and literature review, we present the first empirical study on smart contract maintenance-related concerns. Our study can help smart contract developers better maintain their smart contract-based projects, and we highlight some key future research directions to improve the Ethereum ecosystem.

			
\end{abstract}


\section{Introduction}
\label{sec:introduction}


With the great success of Bitcoin~\citep{bitcoin}, considerable attention has been paid to the emerging concepts of blockchain technology~\citep{blockchain-wiki}. However, the usage scenario of Bitcoin is limited, as the main application of Bitcoin is storing and transferring monetary values~\citep{efanov2018all}. The appearance of Ethereum~\citep{ethereum} at the end of 2015 removed many of the limitations of blockchain-based systems. Ethereum leverages a technology named \textit{smart contracts}, which are Turing-complete programs that run on the blockchain~\citep{Ethereum_yellow_paper}. Blockchain technology gives immutable, self-executed, and decentralized features to these smart contracts. This in turn means that smart contracts \emph{cannot be modified once deployed to the blockchain}, and all of their execution depends on this immutable code. Running these smart contracts across highly distributed servers costs \emph{``gas"}, which in turn costs money. These features ensure the trustworthiness of smart contracts and make the technology attractive to developers and users. By utilizing smart contracts, developers can easily develop Decentralized Applications (\emph{DApps})~\citep{dapp}, which have been applied to different areas, such as IoT~\citep{chen2018iot}, financial~\citep{erc20}, gaming~\citep{cryptokitties}, and data security domain~\citep{velner2017smart}.

Like all computer code, smart contracts may have errors or developers might want to extend their features in the future. However, some features of Ethereum -- like the gas system and smart contract immutability -- make smart contracts much harder to maintain than conventional software~\citep{bosu2019understanding}. Ethereum is a permission-less network and sensitive information -- transactions,  bytecode and balance of smart contracts -- are visible to everyone, and everyone can call the contract by sending transactions~\citep{Ethereum_yellow_paper}. These features increase possible security threats and counter-actions needed. Smart contracts on Ethereum have several other unique characteristics -- the use of the \emph{``gas"} system to fund running of transactions; relatively few patterns and standards for structuring smart contract code; lack of source code available for most deployed smart contracts; and relative lack of tools to check smart contracts for errors, compared to conventional software. All of these features increase the difficulty of smart contract maintenance. 

In software engineering, the term \emph{software maintenance} refers to the modification of a software product after delivery to correct faults and to improve performance or other attributes~\citep{pigoski1996practical}. It is a very broad activity according to the definition of ISO/IEC 14764~\citep{ISO/IEC}. There are four main kinds of maintenance, i.e., adaptive, perfective, corrective, and preventive maintenance. In the context of the four categories of maintenance, the following illustrate the potential impact of such factors on smart contract maintenance:
\begin{itemize}
    \item \textit{\textbf{Adaptive maintenance}} aims to keep software usable in a changed or changing environment. However, the running environment of smart contracts is often unpredictable. For example, smart contracts usually call other contracts. However, the callee contracts might crash and cannot work anymore. Since the callee contracts are immutable, the crash of the callee contract can lead to serious consequences of the caller contract. The unpredictable environment makes it very difficult to conduct adaptive maintenance for smart contracts.
    \item \textit{\textbf{Perfective maintenance}} is used to improve the performance or maintainability by adding new requirements and functionalities newly elicited from users. However, the scalability issues and the gas system of Ethereum make smart contracts difficult to add too many functionalities, else they become very costly to run and unwieldy.
    \item \textit{\textbf{Corrective maintenance}} focuses on fixing discovered bugs and errors in a program. The lack of tools and community support due to the relative newness of smart contracts makes it hard to detect and remove smart contract bugs. 
    \item  \textit{\textbf{Preventive maintenance}} aims to remove latent faults of programs before they become operational faults. For example, a code smell is a characteristic in the source code that possibly indicates a deeper problem~\citep{smellDefinition}. Refactoring the code to remove code smells to increase software robustness is a typical preventive maintenance method. However, due to the immature ecosystem of smart contracts, it is not easy to find appropriate advanced methods to conduct preventive maintenance for smart contracts. 
\end{itemize}


In this paper, we focus on the maintenance-related concerns of post-deployment smart contracts. Unlike traditional programs that can be upgraded directly, \textbf{to maintain a smart contract, developers usually need to redeploy a smart contract and discard the old version.}  Although maintaining smart contracts is not easy, it is still important to find methods to maintain them. For example, in 2016, attackers found the DAO (Decentralized Autonomous Organization) smart contract contains a vulnerability named Reentrancy~\citep{contractDefects, oyente}. This vulnerability was then utilized by attackers and led to the famous DAO attack~\citep{DAOAttack}, which made the DAO lose 3.6 million Ethers (about \$20/Ether before the attack happened). According to recent research~\citep{kalra2018zeus, liu2018reguard}, a similar vulnerability is prevalent in Ethereum smart contracts; all of these contracts can be attacked and lead to financial loss. Thus, it is important to conduct corrective maintenance for these contracts to remove issues like the Reentrancy vulnerability to ensure the contracts are bug-free and robust. 

Many previous works~\citep{zou2019smart, parizi2018empirical, bosu2019understanding, chakraborty2018understanding, li2017survey} conduct empirical studies to investigate the challenges to the entire software development life cycle of smart contracts. This includes smart contract design, programming, security, maintenance, documentation and so on.  However, none focus exclusively on smart contract maintenance. To fill this gap, we provide a comprehensive empirical study on smart contract maintenance based on a systematic literature review that covers 131 smart-contract-related papers selected from a collection of 946 papers to find maintenance-related challenges, and methods for smart contracts. Our study aims to answer the following two key research questions:


\textbf{RQ1: What kinds of maintenance issues will smart contract developers encounter? }

We identify 9 issues related to corrective, adaptive, perfective, and preventive maintenance, and another 4 issues corresponding to the overall maintenance process for smart contracts. These maintenance issues are extracted from previous publications. Since Ethereum and smart contracts are fast-evolving, some results from previous works might be outdated. There might be a gap between academia and industry.  For example, Zhou~\citep{zou2019smart} mentioned that smart contracts miss the support of exception handling, e.g., the \textit{try...catch}. However, Solidity adds the exception handling in v6.0~\citep{Solidity}. To make our results more reliable, we use an online survey to validate our findings. We sent the survey to 1,500 smart contract developers on Github, and received 165 useful responses.  The feedback from the survey can also be a supplement to our findings. We analyze the reasons for smart contract maintenance issues according to the survey results. 

\textbf{RQ2: What are the current maintenance methods for smart contracts? }

To help developers maintain smart contracts, we summarize four kinds of current maintenance methods from 41 publications. 31 publications introduce offline checking methods to help developers maintain smart contracts. They can help maintain smart contracts before they are deployed/redeployed to Ethereum. Seven publications introduced online checking methods, which can help maintain deployed smart contracts by detecting malicious input or automatically upgrading smart contracts. Two previous works suggested developers to use the \textit{Selfdestruct} function to undo contracts when emergencies happen. Another work describes how smart contract can be upgraded by using \textit{DELEGATECALL} instruction.


The main contributions of this paper are: 
\begin{itemize}\setlength{\itemsep}{1pt}
	
	\item To the best of our knowledge, this is the first in-depth empirical study that focuses on the maintenance issues of smart contracts on Ethereum, and we divide the issues into four categories. 
	
	\item Our study identifies the key current maintenance methods used for smart contracts, which gives guidance for smart contract developers to better maintain their contracts. 
	
	\item Our study highlights the limitations and possible future work related to smart contracts on Ethereum. This gives directions for smart contract developers and researchers to develop improved tools and focus future research.

\end{itemize}

The remainder of this paper is organized as follows. In Section 2, we provide background knowledge of smart contracts and Ethereum. In Section 3, we introduce the methodology to conduct the literature reviews and the survey. After that, we present the answers to the two research questions in Sections 4 and 5, respectively. In Section 6, we highlight key threats to validity. We discuss what should be done in the future to improve the Ethereum ecosystem in Section 7 and review related work in Section 8. Finally, we conclude the whole study in Section 9.
\section{Background}
\label{sec:background}

\subsection{Ethereum}
In 2008, the first blockchain-based cryptocurrency named Bitcoin was introduced and demonstrated the enormous potential of blockchain to the world. However, the biggest limitation of Bitcoin is that it only allows users to encode non-Turing-complete scripts to process transactions, which greatly limits its capability. To address this limitation, Ethereum was born at the end of 2015 and brought a revolutionary technology named smart contracts. Nowadays, Ethereum has become the second most popular blockchain system and the most popular platform on which to run smart contracts. Similar to Bitcoin, Ethereum also provides its cryptocurrency and names it as Ether. In Jan. 2018, Ether reached its highest value to \$1389 / Ether~\citep{marketcap}. Unlike Bitcoin, which has a fixed number of coins (21 million in total), 18 million Ethers are created every year~\citep{Ethereum_yellow_paper} (and 72 million Ether were generated at its launch). Currently, two new Ethers are created with each block, and it requires about 14-15s to create a new block; the average Ethereum block size is between 20 to 30 KB, and the biggest Ethereum block size is around 2MB~\citep{ethstats}. Ethereum does not support concurrency, and all transactions need to be executed by all nodes, which leads to a low  throughput of Ethereum. Ethereum only allows about 15 transactions per second on average~\citep{EtherScan}, which has become one of its biggest limitations. At the end of 2017, there is a famous smart-contract-based game named CryptoKitties~\citep{cryptokitties} published in the Ethereum. However, the popularity of the game slowed down all transactions as too many players sent transactions to the Ethereum blockchain. 

\begin{figure}
	\begin{center}
		\includegraphics[width=0.5\textwidth, height=0.14\textheight]{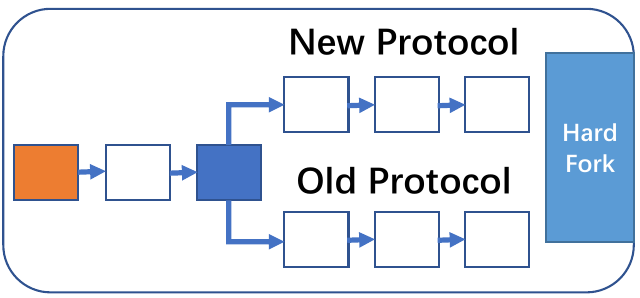} 
		\caption {An Example of Hard Fork. The blue block called divergence block, where the blockchain system updates its protocol. The new protocol for hard fork is not backward-compatible. } 
		\label{Fig:Hard_Fork}
	\end{center}
\end{figure}

\subsection{Hard Fork and Soft Fork}
\label{sec:fork}
Any software or operating system needs periodic upgrades to fix errors or add new functionalities. For the blockchain system, those updates are called a ``fork". There are two kinds of forks, i.e., hard fork and soft fork.

\noindent \textbf{Hard Fork.} Figure~\ref{Fig:Hard_Fork} shows an example of a hard fork. The blockchain system is a decentralized network. All the nodes on the network need to follow the same rules. The set of rules is known as the protocol. In Figure~\ref{Fig:Hard_Fork}, the blue block is called a divergence block, where the blockchain system updates its protocol. When a protocol is updated, and the new protocol is not backwards-compatible. Some nodes on the blockchain do not accept the new protocol, and they choose to use the old version. Thus, the blockchain forks into 2 incompatible blockchains, which run the new and old protocol, respectively. 

\begin{figure}
\begin{center}
	\includegraphics[height=0.14\textheight]{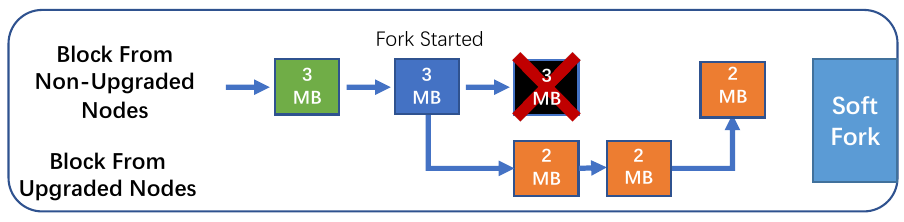} 
	\caption {An Example of Soft Fork. The blue block called divergence block, where the blockchain system updates its protocol. The new protocol for soft fork is backward-compatible. } 
	\label{Fig:Soft_Fork}
\end{center}
\end{figure} 

\noindent \textbf{Soft Fork.} Updates of protocols by soft fork are backwards-compatible.  Nodes that did not upgrade to the new version will still be able to participate in validating and verifying transactions. In this case, there is only one chain on the blockchain by using a soft fork. Noticed that the functionality of a node with the old protocol is also affected. As the example shown in Figure~\ref{Fig:Soft_Fork}, the maximum block size allowed by the old protocol  is 3MB, and the new protocol limits the block size to 2MB. The non-upgraded nodes can still process transactions and push new blocks that are 2MB or less. However, if a non-upgraded node tries to push a block that is greater than 2MB, the upgraded nodes will reject to broadcast the block, which encourages the non-upgraded nodes to update the new protocols.

\subsection{Smart Contracts}
Smart contracts can be regarded as Turing-complete programs that run on the blockchain~\citep{Ethereum_yellow_paper}. They are usually developed in a high-level language, e.g., Solidity, Vyper~\citep{Vyper}. Solidity is the most popular programming language with which to develop smart contracts on Ethereum. Based on the immutable blockchain technology concept, smart contracts cannot be modified once added to the blockchain. Once started, all running of the contract is based on its code. No one can affect it, not even the creator. Ethereum uses EVM (Ethereum Virtual Machine) to execute smart contracts. When developers deploy a smart contract to Ethereum, the contract will be compiled into EVM bytecode, and the bytecode will be stored on the blockchain forever. The only way to remove the bytecode from Ethereum is by using the Selfdestruct function~\citep{Solidity}. There is a unique 40 bytes hexadecimal hash value to identify a contract address. Since Ethereum is a permission-less network; every one can send a transaction to invoke contract functions if they know the function signatures, which includes its function id and parameter types~\citep{Solidity}. Even worse, all the transactions, bytecode, invocation parameters are visible to everyone, which makes smart contracts face major security challenges. 

\subsection{The Gas System}
In Ethereum, transactions are executed by \emph{miners}. To incentivize the execution of smart contracts by miners, transaction senders need to pay an amount of Ether to the miner, which is so-called the \emph{gas mechanism}. For each transaction, the EVM will calculate its gas cost, and the transaction sender is required to define a gas price, e.g., 20 Gwei / gas unit ($1 Ether = 10^9 Gwei$). The final transaction fee is calculated by $gas\_cost  \times gas\_price$. Miners have the right to decide whether or not execute a transaction. Thus, higher gas prices can lead to faster execution, and lower gas prices can lead to a transaction that is never added to a block. According to the \textit{ETH Gas Station}~\citep{ethgasstation}, in May 2020, if the gas price is higher than 40 Gwei, the transaction can be executed within 2 minutes. If the gas price is lower than 25 Gwei, the execution time can exceed half an hour. 

Another function the \emph{``gas"} system is to ensure the execution of smart contracts can be eventually terminated. In Ethereum, the transaction caller is required to set a \textit{gas limit}, which refers to the maximum gas cost of a transaction. If the gas cost of a transaction exceeds the gas limit, the execution will be terminated with an exception thrown by EVM named \textit{out-of-gas error}. 

The gas system ensures the normal running of the Ethereum. However, it also increases the difficulty of  smart contract development, as developers should estimate the maximum gas cost of the contracts. Ethereum block has a maximum size, which limits the amount of data that can be included. The current maximum block size limits the maximum gas limit to 12.5 million gas units~\citep{ethstats}. When the maximum gas cost of a transaction exceeds the 12.5 million, it will be reverted forever. 

\begin{figure}
	\begin{center}
		\includegraphics[width=0.9\textwidth]{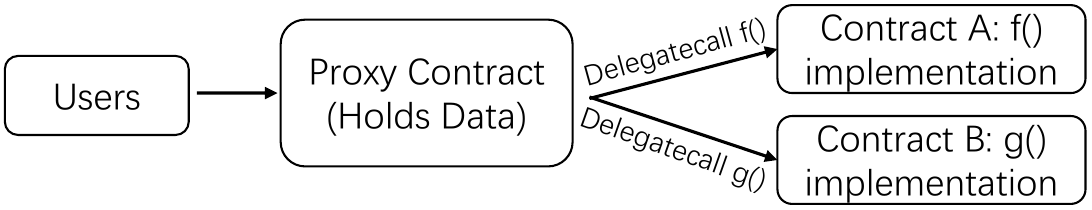} 
		\caption {An Example of the upgradeable contract. } 
		\label{Fig:upgradeable}
	\end{center}
\end{figure}

\subsection{Upgradeable Smart Contracts}

Even though smart contracts cannot be changed once deployed to the blockchain, there is a method to develop ``upgradeable" contracts. Ethereum provides a function named \textit{DelegateCall}, which allows a contract to use code in other contracts, and all storage changes are made in the caller's value. Specifically, \textit{DelegateCall} can be implemented by \textit{addr.delegatecall(bytes memory)}. \textit{addr} is the address of the callee contract (The value of  \textit{addr} can be changed by sending a transaction to the contract). The function selector and input value are encoded as \textit{bytes memory}, and will be sent to the callee contract when \textit{DelegateCall} is executed. Once the execution of the function on the callee contract is finished, the return value will be transferred back to the caller contracts.  When bugs are found at the callee contract, the proxy contract can redirect the \textit{addr} to a new contract. 

Figure~\ref{Fig:upgradeable} is an example of the upgradeable contract, which contains three contracts. The proxy contract holds the data of a contract, and all the storage changes are made in the proxy contract. The proxy contract uses \textit{DelegateCall} to call the functions \textit{f()} and \textit{g()}. These functions are implemented in contract A and B, respectively.  Once errors are found or new functionalities need to be added, contract A and B can be discarded directly. The proxy contract can call the code of the new contract by using \textit{DelegateCall}. Based on this approach, \textit{OpenZeppelin}, a famous smart contract organization, has provided a library~\citep{OpenZeppelin_Upgradeable} to help developers develop upgradeable smart contracts in just a few lines. \textit{EIP 2535}~\citep{EIP2535} (the Diamond Standard) also defines the standard to help developers design upgradeable smart contracts.

\subsection{Software Development and Maintenance}

Software development refers to a set of activities that throughout the entire life cycle of software, which includes the process of designing, creating, deploying and supporting software~\citep{bourque2014guide}. Thus, software maintenance is an important and inevitable part of the software development life cycle. According to previous work~\citep{boehm2005software}, software maintenance can lead to 60\% of software cost. Besides, in many software development models, e.g., Spiral model~\citep{boehm1988spiral}, Agile development~\citep{beck2001manifesto}, it is not easy to split the process of development and maintenance. For example, Agile software development refers to  software development methodologies based on iterative development. In each iteration, new requirements and solutions will be added to improve the software. According to the definition of ISO/IEC 14764~\citep{ISO/IEC}, there are four kinds of software maintenance, i.e., corrective, adaptive, perfective, and preventive maintenance. Among them,  \textit{Perfective maintenance} is used to improve the performance or maintainability by adding new requirements and functionalities newly elicited from users, which is similar to the steps of Agile development. Thus, there are many overlaps between the software maintenance and development.

\subsection{Card Sorting}
Card sorting is a method to organize data into logical groups~\citep{cardsort}. It is widely used to help users understand how a user would organize and structure the data that they think is appropriate. To conduct a card sorting, users first need to identify the key concepts and write them into labeled cards. A card can be everything that helps the discussion, e.g., a piece of paper or a virtual card on a laptop. After that, users are required to group cards into different categories that make sense to them. Due to the low-tech and inexpensive nature of card sorting, it is usually used to design workflow, architecture, category tree, or folksonomy.

There are three kinds of card sorting, i.e., open card sorting, closed card sorting, and hybrid card sorting. Open card sorting is used for organizing data with no predefined groups. Specifically, each card will be clustered into a group with a certain topic or meaning first. If there is no appropriate group, a new group will be generated. All the groups are low-level subcategories and will be evolved into high-level subcategories further. Closed card sorting is used for organizing data with predefined groups. Each card is required to cluster into one of the groups. Hybrid card sorting combines open card sorting and closed card sorting. Hybrid card sorting has predefined groups but allows to create new groups during the process.

\section{Methodology}
\label{sec:methodology}

\begin{figure*}
	\begin{center}
		\includegraphics[width=0.9\textwidth]{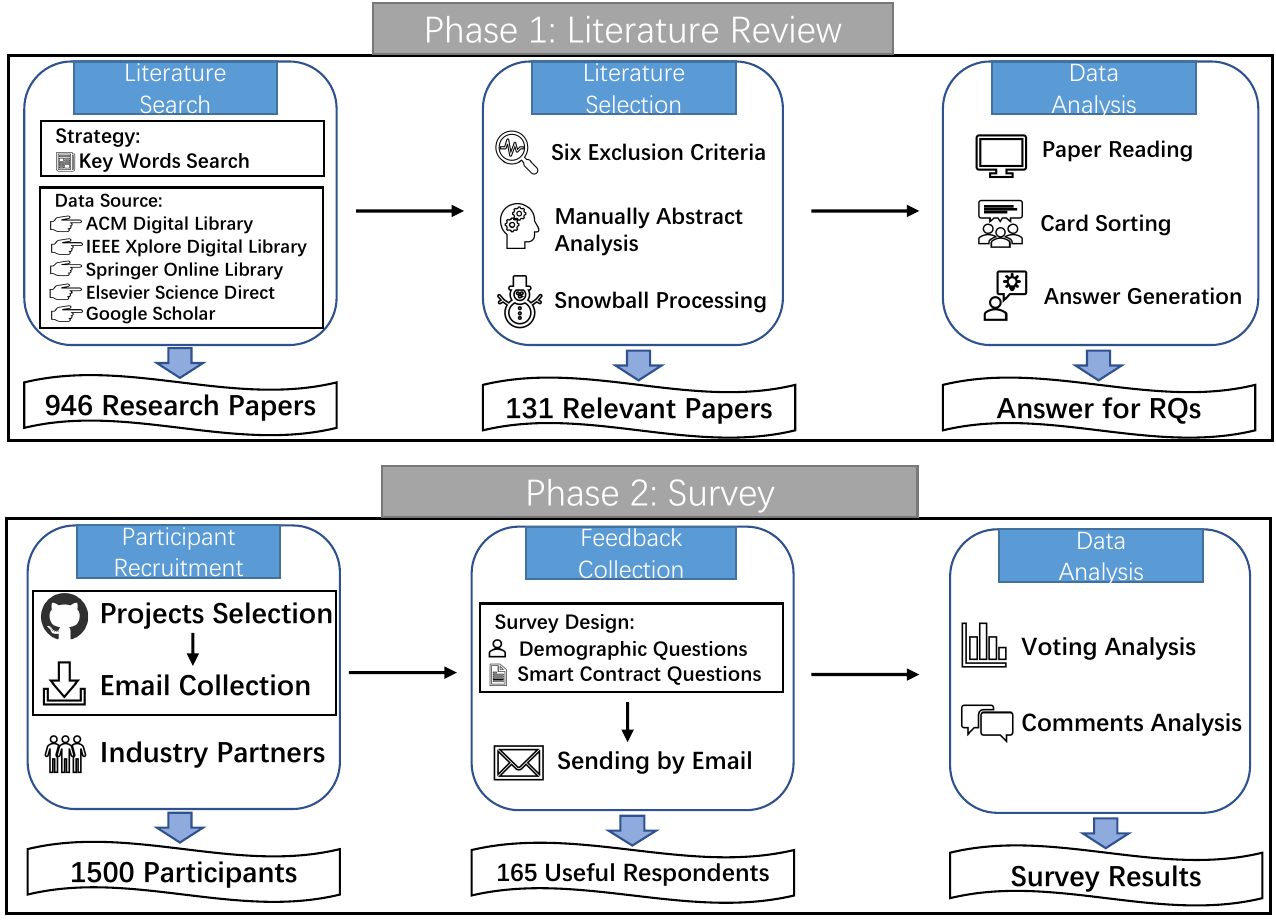} 
		\caption {Overview of methodology design} 
		\label{Fig:overall}
	\end{center}
\end{figure*} 

Figure~\ref{Fig:overall} shows the overview of our methodology, which contains two phases, i.e., literature review and survey. In phase 1, we perform a systematic literature review, which aims to find the answers to research questions from prior smart contract related papers. After obtaining the answers, we use an online survey to validate whether smart contract developers agree with our findings. In the following subsections, we present the detailed steps of our literature review and survey. 

\subsection{Literature Review}
In this paper, we follow the method provided by Kitchenham et al.~\citep{kitchenham2007guidelines} to perform the literature review. There are three steps in phase 1, i.e., literature search, literature selection, and data analysis. 

\subsubsection{Literature Search}
Guided by prior literature reviews~\citep{conoscenti2016blockchain, segura2016survey,  huang2019survey}, we select five search engines, i.e., ACM Digital Library, IEEE Xplore Digital Library, Springer Online Library, Elsevier Science Direct, and   Google Scholar. From these search engines, we can find peer reviewed research papers published in journals, conferences, workshops, and symposia. 

\begin{table}
	\footnotesize
	\caption{Initial Number of Smart Contract Related Research Papers Returned by Each Search Engine}
	\label{tab:libraries}
	\centering
	\begin{tabular}{p{150pt}  r }
		\hline
		\textbf{Search Engine} &  \textbf{Papers}\\
		\hline
		ACM Digital Library &  73\\
		\hline
		IEEE Xplore Digital Library
		& 177\\
		\hline
		Springer Online Library & 54\\
		\hline
		Elsevier Science Direct &  11\\
		\hline
		Google Scholar & 631\\
		\hline
		\textbf{Total} & \textbf{946}\\
		
	\end{tabular}	
\end{table}

We used keyword search to obtain 946 initial smart contract related papers. The detailed numbers of the research papers returned by different search engines are shown in Table~\ref{tab:libraries}. (The duplicated papers are removed.)  All of these 946 research papers contain at least one of the keywords ``smart contracts" , ``smart contract", ``Ethereum", ``blockchain", ``DApps" in their title. Since there are many other blockchain platforms supporting smart contracts, and our focus is Ethereum, all the selected papers should contain the keyword "Ethereum" or ``smart contract" in their abstract. 

\subsubsection{Literature Selection}

Although all the papers that we find in our literature search contain the keywords ``smart contract" or ``Ethereum" in their abstract, some of them are still irrelevant to our study. For example, some research  related to other smart contract platforms might also contain the keyword ``Ethereum" in their abstracts. We applied the following five exclusion criteria to remove irrelevant papers:

\textbf{Exclusion Criteria}

(1) Studies are not written in English.

(2) Master or Ph.D. theses. 

(3) Keynote papers. 

(4) Studies not related to Ethereum.

(5) Studies not related to smart contracts.

In this study we only focus on maintenance-related concerns for post-deployed Ethereum smart contract development issues. Thus, research based on underlying blockchain technology, e.g., consensus algorithms, are excluded. We only focus on the following topics: 

\textbf{Inclusion Topics}

(1) Smart contract empirical studies.

(2) Smart contract security / reliability Analysis. 

(3) Smart contract standards.

(4) Smart contract optimization, e.g., gas optimization. 

(5) Other smart contract technologies, e.g, smart contract generation, decompilers.

To reduce errors, we conducted close card sorting~\citep{cardsort} to check the collected data.  Card sorting is a common method used to evaluate and derive categories from the data~\citep{kim2016emerging}. There are three types of card sort, i.e., open card sort, closed card sort, and hybrid card sorting. Among these three kinds of card sort, closed card sort has predefined categories. We apply closed card sort to select relevant papers, as there only two categories, e.g., relevant or irrelevant. For each card, it has a title (the name of the paper) and description (abstract of the papers). Two experienced researchers with four-year smart contract related experience (including a non-coauthor) carefully read the abstract of the initial 946 research papers independently, and then compare their results after finishing the reading. If there are some differences, they discussed to decide the whether the papers should be excluded. Finally, 112 relevant papers are selected from initial 946 papers. After that, we followed the prior study~\citep{huang2019survey} to conduct a snowballing step to enlarge the paper list. Specifically, we manually checked the references of the identified 112 papers and from these found another 19 papers that satisfy our selection criteria. Thus, we finally selected 131 papers for analysis. The paper list can be found at: \url{https://github.com/Jiachi-Chen/Maintenance}

\begin{figure}
	\begin{center}
		\includegraphics[width=0.75\textwidth]{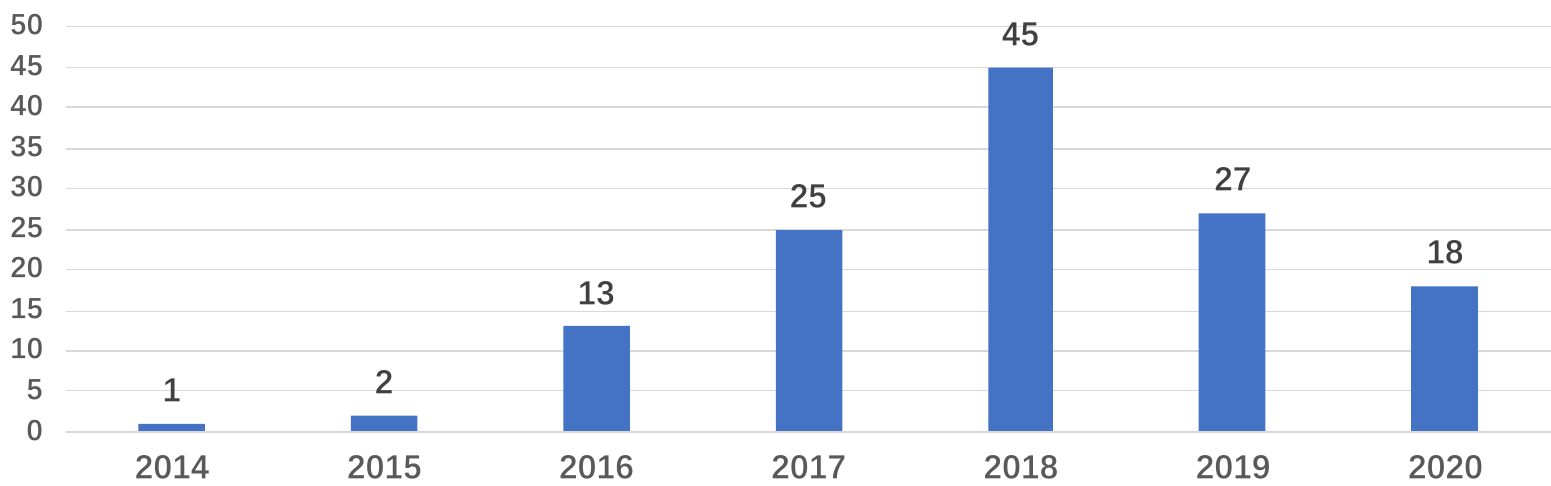} 
		\caption {The number of papers published between 2014 to 2020.} 
		\label{Fig:years}
	\end{center}
\end{figure} 

\begin{figure}
	\begin{center}
		\includegraphics[width=1\textwidth]{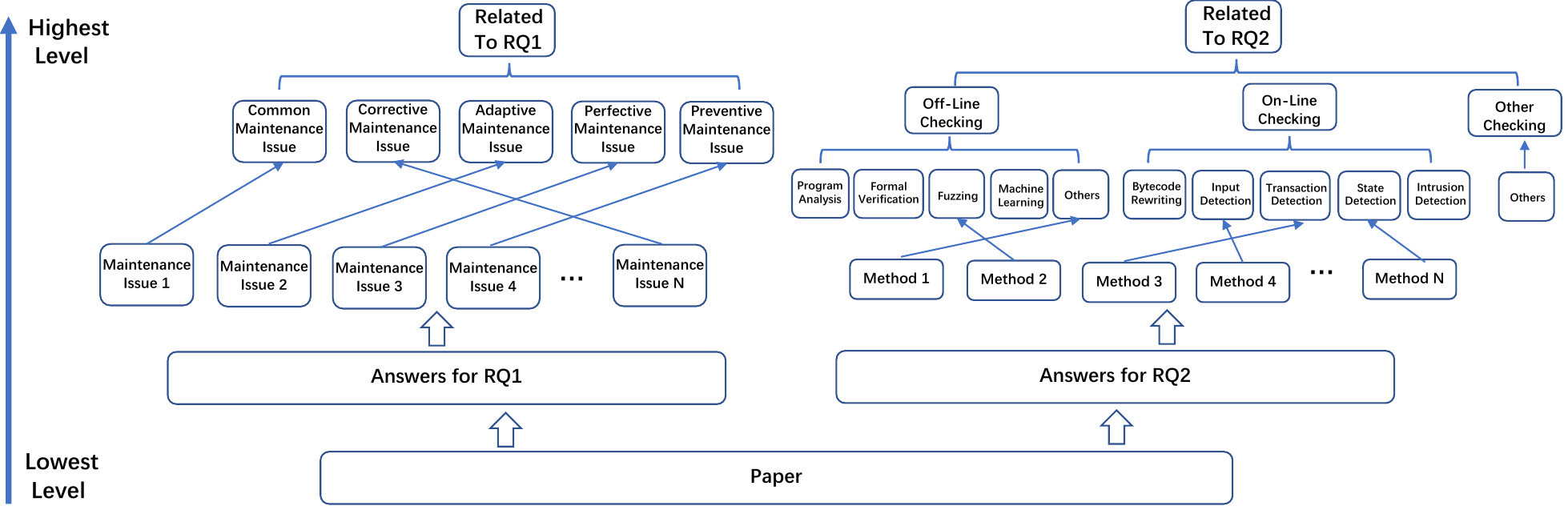} 
		\caption {The steps of open card sorting.} 
		\label{Fig:sorting}
	\end{center}
\end{figure}

\subsubsection{Data Analysis}
The Ethereum proposal was presented in late 2013, and the system went live at the end of 2015. All the 131 selected papers were published between 2014 to 2020 (details see Figure~\ref{Fig:years}), and they were carefully read by the same two researchers. Considering our study aims to find answers with categories being unknown in advance (different kinds of maintenance issues and methods), we decided to adopt an open card sorting approach to help find the answers of these two RQs. The detailed steps are described in Figure~\ref{Fig:sorting}. The two researchers first read the paper carefully and were required to collect the answers to the two RQs shown in Table~\ref{tab:data}, i.e., (1). What are the challenges / issues of smart contract maintenance? (2). What are the used maintenance methods? If we cannot find any answers from a paper, the paper is omitted from our list. For the answers of (1), the data collected from papers were first summarized into detailed maintenance issues. For example, previous works~\citep{chen2018towards, chen2020gaschecker} mentioned that ``..over 90\% of real smart contracts suffer from gas-costly patterns in Ethereum...", which will be summarized into a detailed maintenance issue, i.e., \textit{The Difficulty of Handling the Gas System}. Then, the detailed maintenance issues were clustered according to their maintenance types, e.g., corrective, adaptive, perfective maintenance, and CMI.  For the answers of (2), they were first grouped according to the technique they used, e.g., programming analysis or fuzzing. After that, they will be clustered into a higher level according to their checking types, e.g., off-line / on-line checking. 

\begin{table}
	\scriptsize
	\caption{Data Collection for Each RQ.}
	\label{tab:data}
	\centering
	\begin{tabular}{l  p{300pt} }
		\hline
		\textbf{RQs} &  \textbf{Type of Data We Collected}\\
		\hline
		RQ1 & What are the challenges / issues of smart contract maintenance? The data is classified by corrective, adaptive, perfective, and preventive maintenance. \\
		\hline
		RQ2 & What are the used maintenance methods? e.g., off-line / on-line security checking methods, other methods.\\
		
	\end{tabular}	
\end{table}

\subsection{Survey}

\subsubsection{Survey Design}
Our smart contract developer survey contains three parts, i.e., demographic questions, smart contract maintenance related questions, and suggestion related questions. We follow the previous smart contract related work~\citep{contractDefects} to design the following five demographic questions in our survey. Since our survey is based on Google Form, and Google cannot be accessed in China, we also designed a Chinese version to receive responses from Chinese developers. The translated version was double-checked to ensure consistence with the English version. 

\textbf{Demographics: }
\begin{itemize}\setlength{\itemsep}{1pt}
	
	\item Professional smart contract developer? : Yes / No
	
	\item Involved in open source software development? : Smart Contract Projects only / Traditional Projects Only / Both / None
	
	\item Main role in developing smart contract.
	
	\item Experience in years
	
	\item Current country of residence
	
\end{itemize}	

These questions aim to understand the background and experience of the respondents, which allows us to remove some feedback that we wish to exclude, e.g., feedback provided by very inexperienced respondents. 

In the second part of the survey, we designed 15 questions to help provide answers to the same two research questions that we found from the literature survey. The details of the survey can be found at: \url{https://github.com/Jiachi-Chen/Maintenance}. The list of the questions included in our survey can be found in Table~\ref{tab:survey}. For  questions 1, 3-6, 8-9, 11, we give the participants several choices that are obtained by literature review. Besides, for these questions, we give a textbox to allow participants to write comments. For questions 10 and 12, we follow the previous survey~\citep{contractDefects} to give five scores to participants from score 1 (lowest agreement) to score 5 (highest agreement), and score 3 refers to ``neutral".

\begin{table}
	\scriptsize
	\caption{List of questions included in the survey.}
	\label{tab:survey}
	\centering
	\begin{tabular}{l | p{300pt}}
		\hline
		\textbf{ID} &  \textbf{Question}  \\
		\hline
		Q1 & How do you obtain your required knowledge about smart contracts?\\
		\hline
		Q2 & Do you believe smart contracts have higher security requirements than traditional,  centralized apps, e.g., mobile apps, web apps?   \\
		\hline
		Q3 & How do you test / debug your smart contracts for security and scalability?  \\
		\hline
		Q4 & How do you maintain smart contracts after deployment?  \\
		\hline
		Q5-6 & Have you developed an upgradeable smart contract before? If not, why?  \\
		\hline
		Q7 &  Do you believe smart contracts are harder to maintain than traditional centralized apps, e.g., mobile apps, web apps? Why?\\
		\hline
		Q8 &  What maintenance issues do your smart contracts have?  \\
		\hline
		Q9 &  Which features / limitations of Ethereum can increase the difficulty of maintenance? \\
		\hline		
		Q10 &  Are you satisfied with the current ecosystem for smart contracts, e.g., platforms for sharing data? \\
		\hline
		Q11 &  Have you ever used the code of smart contracts from the following platforms, e.g., Github, Stack Overflow, Etherscan? \\
		\hline		
		Q12 &  Give a score for IDE, testing tools, security audit tools, smart contract explorer, Q\&A site, Comments from Public (Github, DApp Store), community support, Solidity and Ethereum document, respectively. \\
		\hline
		Q13 & Do you think smart contracts are suitable for developing a large scale project? \\
		\hline		
		Q14 & Do you think it is necessary to have an app store like IOS Store for smart contracts?\\
		\hline
		Q15 & Currently, there are many technologies that can improve the security of smart contracts. Do you think it is important to merge them into EVM / Ethereum / IDE?\\
		
	\end{tabular}	
\end{table}

In the third part of the survey, we give a text box to respondents to allow them to give us final comments or questions. 

\subsubsection{Survey Validation}
Guided by Kitchenham et al.~\citep{kitchenham2008personal}, we utilized an anonymous survey~\citep{tyagi1989effects} to collect personal opinions. To increase response rates, we offered a raffle to respondents so that they can choose to leave an email to take part in the raffle to win two \$50 Amazon gift cards. We first sent our survey to our research partners to conduct a small scale test to refine the survey. They were asked to tell us (1) Whether the expressions used in the survey is clear and easy to understand, (2) How many minutes were needed to complete the whole survey. Finally, we modified our survey based on their feedback thus limiting the time for completion of the survey to under 15 minutes. 

\subsubsection{Recruitment of Respondents}
The ideal respondents of our survey are smart contract developers. We aimed to send our survey to Github developers who contributed to smart contract related projects. We first searched for projects on Github by using keywords  ``Smart Contract", ``Ethereum", ``Blockchain", and ranked the projects by the most stars. Then, to increase the response rate and exclude non-smart-contract developers, we manually selected relevant projects by reading the descriptions of the projects. After that, we crawled the emails and names of contributors of the selected projects by using Github Developer API\footnote{https://developer.github.com/v3/}.  We finally obtained 1,500 emails of developers and sent an email to invite them to participate in our survey. We also have some industry partners working in well-known companies, e.g, Alibaba, Facebook, and sent our survey to them (The number of industry partners is 20). Since some developers might not be familiar with ``software maintenance", we inform the concept in the email to reduce the misleading. 

\subsubsection{Data Analysis}
We received a total of 178 valid responses from 32 different countries (The response rate is about 11.87\%), which is a good response number and rate compared to previous smart contract related surveys~\citep{contractDefects, zou2019smart, bosu2019understanding, chakraborty2018understanding, Chen-selfdestruct}. Among these 178 respondents, 13 of them claim that they do not have any experience in smart contract development. Thus, we removed them from our dataset and used the remaining 165 for further analysis. The top three countries in which respondents reside are China (35.76\%), USA (15.15\%) and UK (9.09\%).  The average years of experience in developing smart contracts of our respondents are 2.31 years. Among these respondents, 106 (64.24\%) of them claim their main role is development, 42 (25.45\%) indicate testing/maintenance/evolution, 29 (17.58\%) indicate project management, 6 (3.64\%) indicate risk analysis, 4 (2.42\%) indicate research. (Some respondents have multiple job roles; thus the total number exceeds 165.)

\section{RQ1: What are the maintenance issues of smart contracts?}
\label{sec:rq1}
There are four broad kinds of maintenance, i.e., corrective, adaptive, perfective, and preventive maintenance. In this section, we identify the key maintenance issues for smart contracts considering these four aspects. We also introduce some common maintenance issues (CMI), which appear in all kinds of maintenance. All the findings are obtained by literature reviews (the source are cited), and we give survey results to cross-validate each finding. It should be noted that software maintenance is a very broad activity. Some kind of maintenance, e.g., perfective maintenance also requires developers to develop new functionalities as well as change old. Thus, some of the challenges we discuss can be encountered in both smart contract development and maintenance phases. We use Table~\ref{tab:mapping1} and \ref{tab:mapping2} to help readers better understand the relation between the survey results and the findings collected from literature. The first column of the tables is the survey ID (detailed information can be found at Table~\ref{tab:survey}). In the second column of the table, we highlight the sections that use this survey question to validate related findings, and the related survey results are shown in the third column.

\begin{table}[htbp]
  \centering
  \caption{Part 1 - The mapping between survey questions and related sections with survey results.}
    \begin{tabular}{p{20pt}  p{100pt}   p{180pt}}
    \hline
    Survey ID & Related Section  & Survey Result \bigstrut\\
    \hline
    Q1    & S4.2.1 The Lack of Mature Tools & 52.1\% respondents obtain knowledge from journal and conference papers \bigstrut\\
    \hline
    Q2    & S4.1.2 High Requirement for Security & Smart contracts have higher security requirements (78.18\%) \bigstrut\\
    \hline
    Q3    & S4.2.1 The Lack of Mature Tools & Respondents use program analysis (28.48\%), formal verification(9.09\% ), unit testing(80.61\%), code reviews(73.94\%), functional and integration testing (70.91\%) to test smart contracts \bigstrut\\
    \hline
    Q4    & S4.1.1 No Ideal Deployed Contract Modification Methods & Four methods to maintain a smart contract, and all of them are imperfect. \bigstrut\\
    \hline
    Q5-6  & S4.1.1 No Ideal Deployed Contract Modification Methods (Answer 4 - Developing upgradeable contracts to maintain contracts) & Developing upgradeable contracts can increase development cost and security risks. (32.17\% and 33.04\%) \bigstrut\\
    \hline
    Q7    & S4.1.2 High Requirement for Security & Smart contracts are harder to maintain compared to traditional apps (64.85\%) \bigstrut\\
    \hline
    \multirow{5}[10]{*}{Q8} & S4.2.1 The Lack of Mature Tools & Lack of tools / techniques to audit code. (66.2\%) \bigstrut\\
\cline{2-3}          & S4.4.1 The Scalability Issues & There are not enough useful libraries and APIs (49.7\%); not easy to handle the memory and storage in Solidity programming (38.79\%) \bigstrut\\
\cline{2-3}          & S4.4.2 The Difficulty of Handling the Gas System & It is not easy to handle the gas system when maintaining smart contracts (38.79\%) \bigstrut\\
\cline{2-3}          & S4.5.2 The Lack of High Quality Reference Code & Solidity lacks useful reference code. (38.18\%) \bigstrut\\
\cline{2-3}          & S4.5.3 The Lack of Standards & Ethereum lacks standards (49.7\%) \bigstrut\\
    \hline
    \multirow{5}[10]{*}{Q9} & S4.1.2 High Requirement for Security - Financial Attractiveness & There is more financially attractive for attacking smart contracts (49.09\%) \bigstrut\\
\cline{2-3}          & S4.1.2 High Requirement for Security - Permission-less Network & The permission-less feature could increase the difficulty of maintenance. (55.76\%) \bigstrut\\
\cline{2-3}          & S4.1.3 Low Readability & 89.1\% respondents use the source code of smart contracts (Q11), and 57.14\% of them said the poor readability of smart contracts increases the difficulty of code reuse. \bigstrut\\
\cline{2-3}          & S4.3.1 Unpredictable Fork Problems & Ethereum might add new functions through hard fork, which might affect the currents contracts running on the blockchain. (50.3\%) \bigstrut\\
\cline{2-3}          & S4.3.2 Unpredictable Callee Contracts & It would make their contracts hard to be maintained if the callee contracts crashed or be destructed. (62.42\%) \bigstrut\\

    \end{tabular}%
  \label{tab:mapping1}%
\end{table}%

\begin{table}[htbp]
	\centering
	\caption{Part 2 - The mapping between survey questions and related sections with survey results}
	\begin{tabular}{p{20pt}  p{100pt}   p{180pt}}
		\hline
		Survey ID & Related Section  & Survey Result \bigstrut\\
		\hline
		Q10   & S4.5 Preventive Maintenance Issues & Only 7.88\% and 16.97\% respondents said they are very satisfied or satisfied with the current ecosystems of smart contracts.  \bigstrut\\
		\hline
		Q11   & S4.1.3 Low Readability & 89.1\% respondents claimed that they use the source code of smart contracts from open sourced platforms \bigstrut\\
		\hline
		Q12   & S4.2.2 The Lack of Community Support & The community support receives an average score of 3.03 \bigstrut\\
		\hline
		Q13   & S4.1.1 The Scalability Issues & Only 14.55\% respondents believe smart contracts are suitable for developing a large scale project \bigstrut\\
		\hline
		Q14   & S7.1 Improving the Smart Contract Ecosystem - DApp Store and Comment System. & Having positive opinions about the need for a DApp store like the Android Google Play Store (84.24\%) \bigstrut\\
		\hline
		Q15   & S7.2 Improving Ethereum and Solidity - Merging Cutting-Edge Technologies. & 90.9\% respondents hold positive opinions about merging cutting-edge technologies into the EVM and updated by nodes on Ethereum. \bigstrut\\
		
	\end{tabular}%
	\label{tab:mapping2}%
\end{table}%

\subsection{Common Maintenance Issues}
\subsubsection{No Ideal Deployed Contract Modification Methods}
\label{sec:modification}

Immutability is an important feature of smart contracts, which makes smart contracts distinct from traditional apps in their stability. However, this feature also leads -- intentionally -- to great difficulty for their modification.

From our survey, we received four  answers \footnote{The questions are multi-choice. Thus the sum of each options can exceed 100\%. The same with the other questions.} for the question ``How do you maintain your smart contracts" (Q4 in Table~\ref{tab:survey}). 
The four answers are: 
\begin{enumerate}\setlength{\itemsep}{1pt}
	\item  I never maintained a contract (18.79\%) 
	
	\item  I discard the old contract directly and deploy a new one (39.39\%)
	
	\item I use \textit{Selfdestruct} function to destroy the old contract and deploy a new one (38.79\%) 
	
	\item I develop upgradeable contracts.  (35.76\%). 
	
\end{enumerate}

However, all of these four answers are imperfect and can lead to high financial loss in some situations. 

For answer (1), this method is very inadvisable as some bugs are usually inevitable. Without maintenance, the usefulness life of the programs will be much shortened and attackers can freely attack existing contracts that contain vulnerabilities. 

For answer (2), this method can lead to enormous financial loss for the contract owners, as the Ethers cannot be transferred unless a specific code is included in the contract. Although the contract owners find there is a bug like the reentrancy~\citep{liu2018reguard, rodler2018sereum} in their smart contracts, there was no way to modify the contract, as the contract did not contain a Selfdestruct function and was not develop as an upgradeable contract,  which might lead to an enormous financial loss for the organization. 


For answer (3), adding a \textit{Selfdestruct} function can reduce the financial loss when emergencies happen. Using the DAO attack as an example, if the DAO contract had this function, the DAO organization could use it to destruct the contract and transfer all the Ethers when the attack was detected. After fixing the bugs, they can deploy a new contract, and transfer the Ethers to the new contract. However, this method is still harmful to both contract owners and users in some situations. Our previous work~\citep{Chen-selfdestruct}  investigated the reasons why developers do not add Selfdestruct functions in their contracts. Developer feedback showed the following reasons. \textit{First}, adding a Selfdestruct function also opens an attack vector to the attackers. Thus, developers need to pay more effort to test smart contract security and permissions. The testing can add additional complexity to the development, which can increase the development cost. \textit{Second}, adding a Selfdestruct function can also lead to a trust concern for the smart contract users. This is because many users trust Ethereum because of the immutability of smart contracts. All the execution of the contract depends on its code; even the owner cannot transfer Ethers on the contract balance. This feature is important in financial applications as it ensure the asset safety of contract users. However, the \textit{Selfdestruct} function breaks the immutability of the contracts. It gives power to the contract owners to transfer all the Ethers of the contracts. Thus, this method can lead to the reduction of the number of users of the smart contract using it. \textit{Finally}, the \textit{Selfdestruct} function can also lead to a financial loss in some situations, as the Ethers that were sent to the contract after destroying it will be lost. Thus, this method is still not a perfect method to maintain smart contracts. 

For answer (4), still raises the same trust concern similar to answer (3), as the smart contract immutability features are also be broken. According to our survey (Q5-6 in Table~\ref{tab:survey}), we found that only 29.70\% of the respondents have developed upgradeable smart contracts. There are three reasons why developers do not develop upgradeable contracts. 41.74\% of the respondents claim that they do not know how to develop upgradeable smart contracts. Thus, to develop upgradeable smart contracts, they need to pay a learning cost. 32.17\% and 33.04\% of the respondents said developing upgradeable contracts can increase the development cost and security risks. Thus, this method still incurs a high cost for maintenance. 

To summarize, all of these four methods have disadvantages or limitations, and can lead to a high cost of smart contract maintenance.

\subsubsection{High Requirement for Security}

Unlike traditional programs that can be upgraded directly, developers need to redeploy a new smart contract to the blockchain. Ensuring the security of the contract before redeploying it to the blockchain is important, as each the modification can cost a lot (see~\ref{sec:modification}). According to our survey (Q2 and Q7), 129 (78.18\%) respondents believe smart contracts have higher security requirements. 107 (64.85\%) respondents said smart contracts are harder to maintain compared to traditional apps. The reasons introduced below lead to the high-security requirement of the smart contracts. 

\noindent \textbf{1. The immutability Features}. All the transactions and the code of smart contracts are immutable, which means that developers need to ensure the security of the code and each transaction. Once any bugs are detected, there is no direct way to patch them. Attackers can utilize the errors / bugs to steal Ethers or lock the balance maliciously~\citep{atzei2017survey}. Thus,  immutability raises a high security requirement for the smart contracts.

\noindent \textbf{2. Financial Attractiveness}. Financial profit is an important motivation for attackers. According to our survey (Q9), about 81 (49.09\%)  respondents believe that there is more financially attractive for attacking smart contracts compared to traditional software, thus leading to more attack~\citep{torres2019art}. Since many contracts hold Ethers, attackers can earn profits through their attacks. Even worse, the sensitive information of smart contracts are visible to anyone, e.g., bytecode, Ethers on the balance. Attackers can launch precision strikes to the vulnerable contracts.  Thus, developers need to pay more efforts to ensure the security of smart contracts. 


\noindent \textbf{3. Permission-less Network}. Ethereum smart contracts run on a permission-less network; everyone can execute the smart contracts by sending a transaction. 92 (55.76\%) respondents (Q9) mentioned that the permission-less feature could increase the difficulty of the maintenance. They need to pay more effort to test the permission of the contracts. Previous work~\citep{Chen-selfdestruct} introduced a security issue named \textit{Limits of Permissions}.  Some contracts do not check the permission of their sensitive functions. Attackers can utilize the vulnerabilities of the permission check to steal Ethers.  


\subsubsection{Low Readability}
\label{sec:cmi4}
Readability is important to help developers understand the smart contracts and maintain their smart contracts~\citep{zou2019smart}. According to our survey, 147 (89.1\%) respondents (Q11) claim that they use the source code of other smart contracts from open sourced platforms, e.g., Etherscan, Github to help author and maintain their smart contracts. 57.14\% of the respondents (Q9) also said the poor readability of smart contracts increases the difficulty of code reuse.  Making smart contracts readable is a challenge, as developers need to balance the readability with gas consumption. For example, optimizing code is a common method to reduce gas consumption. The more gas-efficient code usually corresponds to shorter code. However, this shorter code can lead to poorer readability. 

\subsubsection{The Lack of Experienced Developers and Researchers.}
Experienced developers and researchers are the main  inventors of new advanced SE methods to address the limitation of smart contracts, e.g., developing tools, improving ecosystem. However, our survey results and  literature review shows that less experienced people programming in Ethereum compared to traditional development. 

Ethereum is a young system, which was published in 2016. The most experienced developers and researchers of the respondents of the survey have 4 years experience (22 respondents) in smart contracts development, the minimum, average, and median numbers  are 0.2, 2.31, and 2.5 years, respectively. Compared to the experiences of the respondents (including developers and researchers) of previous works, e.g.,  in machine learning~\citep{wan2019does} (min: 3, max: 16, median: 6, avg: 7.6 years),  in desktop software development~\citep{wan2018perceptions} (min: 3, max: 12, avg: 6.5 years), the smart contract developers and researchers seem less experienced. 

\subsection{Corrective Maintenance Issues}
It is not easy to discover all potential bugs before deploying smart contracts to the blockchain. Some bugs / errors of the contracts might be exposed to the public under certain situations. Corrective maintenance is the modification of a smart contract after deployment to the blockchain to correct discovered bugs / errors. Diagnosing errors of smart contracts is the major task in corrective maintenance. However, it is painful and difficult to diagnose errors in a smart contract. According to our survey, 96 (66.2\%) respondents (Q8) complain that debugging and testing is not easy. There are two main reasons that lead to the difficulty of the diagnosing errors, i.e., the lack of mature tools and community support. 

\subsubsection{The Lack of Mature Tools}

Many previous works~\citep{zou2019smart, norvill2017automated, bosu2019understanding} mentioned that smart contract development lacks appropriate tools / techniques to verify code correctness. Thus, it is not easy to fix bugs in smart contracts. A similar theme is also received in our survey. 96 (66.2\%) respondents (Q8) claim that they cannot find useful tools to debug / test / audit their contracts. However, with the development of smart contract ecosystems, a large number of tools have been developed. For example, tools based on static analysis~\citep{oyente, liu2018reguard, tikhomirov2018smartcheck} and formal verification~\citep{bhargavan2016formal, bigi2015validation, hildenbrandt2018kevm} have been proposed. Some tools have excellent performance and speed in detecting common security issues. Thus, ``lack of tools" seems to be addressed with the effort of researchers and developers. There is a gap between academia and industry, as many tools developed in academia are not yet known about and used in industry.

To find the reason, we asked how developers obtain their required knowledge about smart contracts. The Solidity documentation, blogs, and Q\&A website are the top three most popular sources to acquire knowledge; the numbers are 149 (90.3\%), 114 (69.1\%), and 88 (53.3\%), respectively (Q1). The state-of-art tools usually published in academic journal and conference papers, and 86 (52.1\%) respondents (Q1) said journal and conference papers are an important approach to require knowledge. Thus, the methods to require knowledge is not the main reason why developers think that there are not enough tools.

We also investigated the usage conditions for different kinds of tools and how developers test their contracts. We found that only 47 (28.48\%) and 15 (9.09\%) respondents (Q3) use static analysis tools and formal verification tools to test their smart contracts. Unit testing, code reviews, functional and integration testing are still the most popular methods to test smart contracts. About 80.61\%, 73.94\%, and 70.91\% of respondents (Q3) choose these methods to test their contracts. Developer comments said that ``although there are many tools that can be chosen, most of them are hard to use and not user friendly". Thus, although there is a large number of tools that have been developed, developers still complain there are only a few tools they think can be used in practice. 

\subsubsection{The Lack of Community Support}
Community support is a primary source of knowledge for blockchain software projects~\citep{chakraborty2018understanding}. Community support consists of many parts. For example, when developers encounter technical problems, a Q\&A website such as Stack Overflow is an important source to help them address the problems. Developers can open source their projects to Github. Other developers can submit issue reports to help them polish the projects. The App store is also an important place to receive reviews.  Reviews might contain feature requests, user feedbacks, issue reports that can help developers upgrade their software. 

However, community support is not enough for smart contract developers. Previous works~\citep{zou2019smart, hegedHus2019towards} found that smart contract developers lack community support as the blockchain technology is new and there are not enough smart contract developers to answer their questions. Since more and more developers take part in smart contract development, we used our survey to investigate whether community support is still lacking in Ethereum. 

In our survey, we asked respondents to give a score for the community support (Q12). Score 1 refers to `very unsatisfied', 3 refers to `neutrality', and 5 means `very satisfied'. The community support receives an average score of 3.03, while the score for other comparative items e.g., Solidity document, and Smart contract Explorer receive scores of 3.53 and 3.52, respectively. Thus developers still believe that community support is not sufficient compared to other resources. Surprisingly, the score for the Q\&A website, e.g, Stack Overflow, is 3.43, which can show that the Q\&A website is not the culprit for the lack of community support. We found that the score for the ``Comments from public (E.g., DApp, Github)" is only 2.57, which is the lowest score among all the comparative items. 




Previous works~\citep{zou2019smart, hegedHus2019towards} claimed that smart contract developers lack  community support because there are not enough smart contract developers to answer technical questions. However, our survey shows a different answer. The culprit for the lack of community support is not the Q\&A website, but the comments from the public, e.g., issue reports from Github, comments from App Store. 


\subsection{Adaptive Maintenance Issues}
Adaptive maintenance aims to keep a software product usable in a changed or changing environment. In traditional software, the environment changes are usually reflected in the upgrading of the operating systems, the hardware, or software, e.g., database. Conducting adaptive maintenance for the traditional environment changing is not difficult, as these kinds of environment changes are predictable. For example, the updated operating systems usually will give a specific date and detailed API documents. 

However, the environment of smart contracts is more unpredictable. In this subsection, we highlight two challenges, which makes it is not easy to conduct adaptive maintenance for smart contracts.

\subsubsection{Unpredictable Fork Problems}
Ethereum uses soft forks and hard forks (See Section~\ref{sec:fork}) to update the blockchain system. Some forks are planned, while some are controversial unpredictable forks, which might result in smart contract maintenance needs.

In a planned fork, developers are informed in advance, and they usually do not need to update the code of smart contracts. For example, in 2017, a hard fork named ``Byzantium" of Ethereum added a `REVERT' opcode, which permits error handling without consuming all gas~\citep{EIP140}. The function \textit{revert()} in smart contract code will refer to the new opcode automatically. Thus, the planned forks are more likely to be accepted by miners and developers. 

However, unplanned forks are also common in Ethereum, which can increase the difficulty of smart contract maintenance. The first unplanned fork happened in July 2016 and was the result of the DAO attack~\citep{DAOAttack}. The DAO attack made the DAO (Decentralized Autonomous Organization)  lose 3.6 million Ethers. To retrieve the loss, the DAO appealed for a hard fork. The hard fork reversed all the transactions to the block before the attack. This hard fork is controversial, as many miners believe it breaks the law of Ethereum. The opposition miners did not take part in the fork, and a new blockchain was generated, named Ethereum Classic (ETC)~\citep{ETC}. After the hard fork, both ETC and Ethereum contain the same smart contracts. Thus, which contracts to maintain might be a problem for some developers. The same situation also happened to their callee contracts. For example, contract A has two callee contracts, i.e., contract B and C. Unfortunately, contract B chooses to maintain the contract on ETC, while contract C chooses to maintain the contract on Ethereum. Thus, contract A will always have a unmaintained callee contract. 

In Oct. 2016 and Nov. 2016, two unpredictable hard forks were launched to address different problems that have arisen from the DoS attacks. These two hard forks named ``EIP-150 Hard Fork"~\citep{eip150} and ``Spurious Dragon"~\citep{sd-hardfork}, respectively. In ``EIP-150 Hard Fork", Ethereum increased the gas cost of every type of call from 40 to 700 unit. The ``Spurious Dragon" also increases the gas cost of the ``\textit{EXP}" opcode. This increased gas cost might increase the risk of ``out-of-gas error". Thus, some contracts need to refactor their code to handle these gas cost changes. 

According to our survey, 83 (50.30\%) respondents (Q9) are afraid that the forks of Ethereum might result in various potential problems for their smart contracts. Moreover, the unpredictable forks make it difficult for developers to perform adaptive maintenance.

\subsubsection{Unpredictable Callee Contracts}
\label{sec:callee}
Ethereum is a permission-less network; everyone can call the function of the smart contract by sending a transaction. Michael et al.~\citep{frowis2017code} investigated the call relations of smart contracts on Ethereum by checking the hard code address on their bytecode. They found that it is very common for smart contracts to call each other in Ethereum. However, they also found that many callee contracts on Ethereum contain vulnerabilities. These vulnerabilities might lead to the crash and make the contracts cannot work anymore. Beside, many callee contracts also contain \textit{selfdestruct} function, which allow their contract owners to destruct the contracts. Once a contract is destructed, the contract cannot be called anymore, and all the Ethers sent to the destructed contract will be locked forever. 




According to our survey (Q9), 103 (62.42\%) respondents said it would make their contracts hard to be maintained if the callee contracts crashed or be destructed.

\subsection{Perfective Maintenance Issues}
\label{sec:perfective}
As long-lived software~\citep{lohrmaintenance}, users are likely to elicit new requirements during the entire smart contract life cycle. Thus, adding additional functionalities, performance enhancement, and efficiency and maintainability improvements for smart contracts are necessary to respond to the new requirements. This is called the perfective maintenance of smart contracts. Thus, there is an overlap between perfective maintenance issues with development issues, as some new functionalities are required to be developed during this maintenance process. 

However, due to the scalability issues of Solidity and EVM, it is not easy to add too many functionalities to  smart contract-based projects. The Gas system also increases the difficulty of perfective maintenance. Due to these issues, we find that only 24 (14.55\%) of the respondents (Q13) of our survey believe smart contracts are suitable for developing a large scale project.

\subsubsection{The Scalability Issues}
\label{sec:scalability}
\noindent \textbf{Solidity.} Solidity is the most popular programming language for smart contract development, which is an object-oriented language and a bit like JavaScript. However, the grammar of Solidity is too simple to support large projects, which lead to the scalability issues of smart contracts~\citep{zou2019smart}. First, 82 (49.70\%) respondents (Q8) to our survey said there are not enough useful libraries and APIs. Thus, developers need to develop various kinds of APIs and libraries which increases the difficulty of implementing new requirements. Besides, 62 (37.58\%) and 64 (38.79\%) respondents (Q8) also said it is also not easy to handle the memory and storage in Solidity programming, respectively. For example, Solidity only allows creating 16 local variables in a function. Thus, developers have to use storage variables instead of  local variables.  Peter et al.~\citep{hegedHus2019towards} investigated more than 40,000 smart contracts on Ethereum using 16 metrics, e.g., LOC, nesting level. They found the smart contracts are neither overly complex nor coupled much, and do not rely heavily on inheritance. Their results also prove that real-world smart contracts are small-scale programs and do not contain too many functionalities. 

\noindent \textbf{EVM.} The Ethereum Virtual Machine (EVM) is the runtime environment for smart contracts in Ethereum. Some features of EVM make it scale poorly to support large-scale projects. First, EVM does not support multi-thread execution, which makes the execution of smart contracts inefficient. In some large-scale projects, it is important to execute multiple functionalities in parallel to increase  execution speed~\citep{zou2019smart}. Second, EVM limits the maximum size of stack to 1024 items with 256 bits for each item.  The limited stack sizes can easily lead to vulnerabilities and increase the difficulty of developing complex applications~\citep{oyente}. Finally, EVM uses a key-value store, which is a very simplistic database and can lead to low efficiency~\citep{grech2019gigahorse}.

\noindent \textbf{Ethereum. } Ethereum does not support concurrency. To construct the blockchain and ensure security, each node on Ethereum stores the entire transaction history and current state of Ethereum, e.g., account balance, contract variables. Thus, all transactions must be executed and verified by all the nodes. This mechanism makes Ethereum support only around 15 transactions per second, leading to serious scalability issues of smart contract applications. ~\citep{bez2019scalability} 


\subsubsection{The Difficulty of Handling the Gas System}
Ethereum adopts a unique gas system to execute the computational cost of each transaction. The gas system ensures the normal running of the Ethereum system, e.g., giving rewards for miners, avoiding DoS Attack. However, this gas system is also not easy to use, especially when the scale of the project becomes larger. According to our survey, 64 (38.79\%) respondents (Q8) claim that it is not easy to handle the gas system when maintaining their smart contracts. 

First, users need to pay Ethers for the gas cost, and the gas cost depends on the computational cost of the code. Thus, it is important for developers to reduce the gas cost. As we discussed in Section~\ref{sec:cmi4}, there is a trade-off between the gas cost and the readability, and readability is very important for maintenance and large-scale projects. According to previous works~\citep{chen2018towards, chen2020gaschecker}, over 90\% of real smart contracts suffer from gas-costly patterns in Ethereum. However, fixing these gas-costly patterns reduce the readability of smart contracts. 


\subsection{Preventive Maintenance Issues}
Preventive maintenance aims to lessen the likelihood of a sudden breakdown of the programs~\citep{tai1998board}. Guided by advanced software engineering theories, preventive maintenance usually involves some form of redesign or refactor of a smart contract to remove latent faults / errors/ bugs. For example, a code smell is not a bug but are any characteristics in the source code that possibly indicates a deeper problem~\citep{smellDefinition}. Refactoring the code to remove code smells in software to increase its robustness is a typical preventive maintenance method. However, due to the immature ecosystem of smart contracts, it is not easy to find appropriate advanced software engineering (SE) methods, e.g., code smells for smart contracts, to perform preventive maintenance. According to our survey (Q15), only 13 (7.88\%) and 28 (16.97\%) respondents said they are very satisfied or satisfied with the current ecosystems of smart contracts. 

\subsubsection{The Lack of Advanced SE Approach and Research Data}
During our literature review, we found that there are only a small number of works that propose advanced SE methods to help conduct the preventive maintenance of smart contracts. Most of these works aim to improve the reliability of smart contracts, e.g., security check tools (detailed introduced in Section~\ref{sec:rq2}). Compared to traditional software, the maintenance methods of smart contracts to remove latent errors are much less, e.g., code smell removal~\citep{fontana2016comparing}, bug prediction~\citep{giger2012method},  self-admitted technical debt determination~\citep{yan2018automating}.  The lack of research data is an important issue. 

In traditional software maintenance, a large number of MSR (Mining Software Repository) methods have been developed to help conduct preventive maintenance. For example, history bug reports can be utilized to predict whether a source code file contains latent errors ~\citep{zhang2019where2change}. User reviewers can provide feature requests to help developers improve the programs~\citep{maalej2015bug, grano2017android}. Comments in source code can be used to detect self-admitted technical debate, which can be used to signal future errors ~\citep{yan2018automating}. Privacy policies, Stack Overflow (SO) posts, error messages, and commit messages are wildly used to help maintain traditional apps. These methods are not difficult to be applied to smart contract projects. However, the lack of related research data makes it is not easy to develop advanced SE methods for smart contracts.

\subsubsection{The Lack of High Quality Reference Code} 
High-quality reference source code can be a good example when developers conduct preventive maintenance. However, the qualities of open-source smart contracts are poor in Ethereum, and 63 (38.18\%) respondents (Q8) of our survey mentioned that Solidity lacks useful reference code.  

He et al.~\citep{he2019characterizing} found that the copy-paste vulnerabilities were prevalent in Ethereum, and over 96\% of smart contracts have duplicates, which means the ecosystem of smart contracts on Ethereum is highly homogeneous. Among these contracts, 9.7\% of them have similar vulnerabilities. Similar findings are reported by Kiffer et al.~\citep{kiffer2018analyzing}; they investigated 1.2 million contracts, and they can be reduced to 5,877 contract ``clusters" that have highly-similar bytecode. The highly homogeneous nature of smart contracts show that only a limited number of contracts can be referenced during maintenance and development. 

Kiffer et al.~\citep{kiffer2018analyzing} also found that more than 60\% of smart contracts are never actually called. Most of these contracts are useless and hard to be reused. Similar findings were also reported by~\citep{di2019mayflies}. They analyzed the bytecode of smart contracts on Ethereum and found 44,883 are useless and hard to be reused. Only 0.6\% of the contracts have more than 1,000 transactions, while most of the active contracts are similar ERC20 contracts~\citep{erc20}, which are used to make tokens. Thus, the active contracts also cannot provide too much reference value. 

Heged{\H{u}}s et al.~\citep{hegedHus2019towards} analyzed more than 40 thousand Solidity source files. They found that the open sourced smart contract code either quite well-commented or not commented at all. Without comments in the source code, it is not easy for developers to understand and reuse the reference code. 

\subsubsection{The Lack of Standards}

Standards can give guidance for developers to increase the maintainability and reliability of their smart contracts, which is the main motivation for preventive maintenance.  For example, the ERC 20~\citep{erc20} standard defines some rules for token-related contracts. The rules contain 9 functions (3 are optional) and 2 events. This standard allows any tokens on Ethereum to be re-used by other applications, e.g., wallets, decentralized exchanges. At the end of 2017, the CryptoKitties~\citep{cryptokitties} was published and swept the globe. To help other developers develop similar applications, ERC 721 was published in Jan. 2018. ERC 721 is a standard that describes how to build non-fungible tokens (NFTs) on Ethereum, and a NFT is a unit of data on blockchain that represents an unique digital asset, e.g., a photo or a game. Developers can conduct preventive maintenance to make their contracts follow the ERC 721 standard. Thus, their applications can much more easily interact with other similar applications.

However, there are only limited numbers of smart contract related standards~\citep{EIP}. According to our survey (Q8), 82 (49.70\%) respondents said Ethereum lacks standards, which increases the difficulty of the maintenance of smart contracts.

\section{RQ2: What  are  the  current  maintenance  methods  for  smart contracts?}
\label{sec:rq2}
We discuss answers found for our second Research Question, and introduce the current smart contract maintenance methods identified from 41 analysed research papers. 

\begin{figure}
	\begin{center}
		\includegraphics[width=0.5\textwidth]{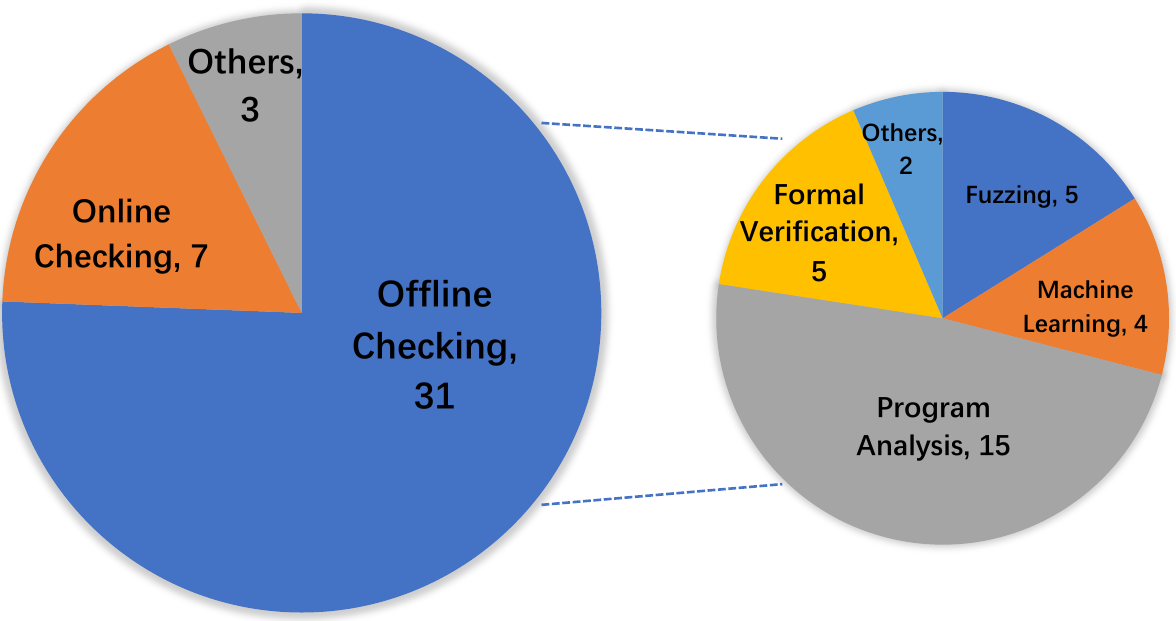} 
		\caption {Distribution of Maintenance Methods} 
		\label{Fig:distribution}
	\end{center}
\end{figure}

\subsection{Distribution}
Among our 131 smart contract selected papers, 41 papers proposed methods that can be used to maintain smart contracts. Unlike traditional software where programs can be upgraded directly, smart contracts need to redeploy new versions to the blockchain and discard old versions. Most maintenance methods check security issues of smart contracts before redeploying them to the blockchain, which are so-called \emph{offline checking methods}. There are 31 papers related to this topic. 7 research papers propose methods that can help maintain a deployed smart contracts. This kind of method is called an \emph{online checking method}. The final three papers introduce a method that uses \textit{DELEGATECALL} to upgrade a smart contract, and a method that redeploys smart contracts by using \textit{Selfdestruct} function, respectively. The distribution of these methods is shown in Figure~\ref{Fig:distribution}.

\begin{table*}[htbp]
	\centering
	\footnotesize
	\caption{Literature of Offline Checking Methods}
	\begin{tabular}{p{70pt}| p{210pt} | p{20pt}}
		\hline
		Category & Name of Publications & Years \bigstrut\\
		\hline
		\multirow{16}[32]{*}{Program Analysis} & OSIRIS: Hunting for Integer Bugs in Ethereum Smart Contracts~\citep{torres2018osiris} & 2018  \bigstrut\\
		\cline{2-3}          & The art of the scam: Demystifying honeypots in Ethereum smart contracts~\citep{torres2019art} & 2019  \bigstrut\\
		\cline{2-3}          & Security Assurance for Smart Contract~\citep{zhou2018security} & 2018  \bigstrut\\
		\cline{2-3}          & Vandal: A Scalable Security Analysis Framework for Smart Contracts~\citep{brent2018vandal} & 2018 \bigstrut\\
		\cline{2-3}          & MadMax: surviving out-of-gas conditions in Ethereum smart contracts~\citep{grech2018madmax} & 2018  \bigstrut\\
		\cline{2-3}          & Finding The Greedy, Prodigal, and Suicidal Contracts at Scale~\citep{nikolic2018finding} & 2018  \bigstrut\\
		\cline{2-3}          & sCompile: Critical Path Identification and Analysis for Smart Contracts~\citep{chang2019scompile} & 2019  \bigstrut\\
		\cline{2-3}          & teether: Gnawing at Ethereum to Automatically Exploit Smart Contracts~\citep{krupp2018teether} & 2018  \bigstrut\\
		\cline{2-3}          & Making Smart Contracts Smarter~\citep{oyente} & 2016  \bigstrut\\
		\cline{2-3}          & Manticore: A User-Friendly Symbolic Execution Framework for Binaries and Smart Contract~\citep{mossberg2019manticore} & 2019  \bigstrut\\
		\cline{2-3}          & SmartCheck: Static Analysis of Ethereum Smart Contracts~\citep{tikhomirov2018smartcheck} & 2018 \bigstrut\\
		\cline{2-3}          & TokenScope: Automatically Detecting Inconsistent Behaviors of Cryptocurrency Tokens in Ethereum~\citep{chen2019tokenscope} & 2019  \bigstrut\\
		\cline{2-3}          & Towards saving money in using smart contracts~\citep{chen2018towards} & 2018 \bigstrut\\
		\cline{2-3}          & GasChecker: Scalable Analysis for Discovering Gas-Inefficient Smart Contracts~\citep{chen2020gaschecker} & 2020 \bigstrut\\
		\cline{2-3}          & Securify: Practical Security Analysis of Smart Contracts~\citep{Securify} & 2018 \bigstrut\\
		\hline
		\multirow{5}[8]{*}{Formal Verification} & Formal Verification of Smart Contracts~\citep{bhargavan2016formal} & 2016  \bigstrut\\
		\cline{2-3}          & A formal verification tool for Ethereum VM bytecode~\citep{park2018formal} & 2018 \bigstrut\\
		\cline{2-3}          & Kevm: A complete formal semantics of the Ethereum virtual machine~\citep{hildenbrandt2018kevm} & 2018  \bigstrut\\
		\cline{2-3}          & Towards verifying Ethereum smart contract bytecode in Isabelle/HOL~\citep{amani2018towards} & 2018  \bigstrut\\
		\cline{2-3}          & ZEUS: Analyzing Safety of Smart Contracts~\citep{kalra2018zeus} & 2018  \bigstrut\\
		\hline
		\multirow{5}[8]{*}{Fuzzing} & ContractFuzzer: fuzzing smart contracts for vulnerability detection~\citep{jiang2018contractfuzzer} & 2018 \bigstrut\\
		\cline{2-3}          & ReGuard: Finding Reentrancy Bugs in Smart Contracts~\citep{liu2018reguard} & 2018 \bigstrut\\
		\cline{2-3}          & EVMFuzz: Differential Fuzz Testing of Ethereum Virtual Machine~\citep{fu2019evmfuzz} & 2019 \bigstrut\\
		\cline{2-3}          & sFuzz: An Efficient Adaptive Fuzzer for Solidity Smart Contracts~\citep{nguyen2020sfuzz} & 2020 \bigstrut\\
		\cline{2-3}          & Exploiting the Laws of Order in Smart Contracts~\citep{kolluri2019exploiting} & 2019  \bigstrut\\
		\hline

		\multirow{4}[8]{*}{Machine Learning} & S-gram: Towards Semantic-Aware Security Auditing for Ethereum Smart Contracts~\citep{liu2018s} & 2018  \bigstrut\\
		\cline{2-3}          & Hunting the Ethereum Smart Contract: Color-inspired Inspection of Potential Attacks~\citep{huang2018hunting} & 2018  \bigstrut\\
		\cline{2-3}          & Towards Safer Smart Contracts: A Sequence Learning Approach to Detecting Security Threats~\citep{tann2018towards} & 2019 \bigstrut\\
		\cline{2-3}          & Checking Smart Contracts with Structural Code Embedding~\citep{gao2020checking} & 2020  \bigstrut\\
		\hline
		\multirow{2}[6]{*}{Others} & Designing Secure Ethereum Smart Contracts: A Finite State Machine Based Approach~\citep{mavridou2018designing} & 2018 \bigstrut\\
		\cline{2-3}          & Mutation Testing for Ethereum Smart Contract~\citep{li2019musc} & 2019  \bigstrut\\
		
	\end{tabular}%
	\label{tab:offline}%
\end{table*}%

\begin{table*}
	\caption{Literatures of Online Checking Methods.}
	\label{tab:online}
	\centering
	\small
	\begin{tabular}{p{70pt} | p{210pt} | p{20pt} }
		\hline
		Methodology & Name of Publications &  Years\\
		\hline
		Bytecode Rewriting & Smart Contract Defense through Bytecode Rewriting~\citep{ayoade2019smart} & 2019 \\
		\hline
		Bytecode Rewriting & Monitoring smart contracts: ContractLarva and open challenges beyond~\citep{azzopardi2018monitoring} & 2018 \\
		\hline
		Input Detection & Town Crier: An Authenticated Data Feed for Smart Contracts~\citep{zhang2016town} & 2016\\
		\hline
		Input Detection & FSFC: An input filter-based secure framework for smart contract~\citep{wang2020fsfc} & 2020 \\
		\hline
		Transactions Detection & {\AE}GIS: Smart Shielding of Smart Contracts~\citep{ferreira2019aegis} & 2019\\
		\hline
		Transactions Detection & VULTRON: Catching Vulnerable Smart Contracts Once and for All~\citep{wang2019vultron} & 2019 \\
		\hline
		State Detection & Sereum: Protecting Existing Smart Contracts Against Re-Entrancy Attacks~\citep{rodler2018sereum} & 2018 \\
		\hline
		Intrusion Detection & ContractGuard: Defend Ethereum Smart Contracts with Embedded Intrusion Detection~\citep{wang2019contractguard} & 2019 \\
		
	\end{tabular}	
\end{table*}

\subsection{Offline Checking Methods}
Table~\ref{tab:offline} summarises the 31 publications which use offline checking methods to help maintain smart contracts. Developers can use the proposed methods to check for security vulnerabilities to help them to maintain smart contracts. For example, using the proposed methods to locate bugs during corrective maintenance, and checking for vulnerabilities of the update versions before redeploying them to Ethereum. We divide the methods presented in these papers into five categories -- program analysis, fuzzing, formal verification, machine learning, and others. In the following subsections, we discuss some key examples. 

\subsubsection{Program Analysis}
\noindent \textbf{CFG (Control Flow Graph) Based Tools.} In 2016, Luu et al.~\citep{oyente} identified four kinds of new security issues of smart contracts and proposed the first tool, named \textit{Oyente}, to detect them through Ethereum bytecode. Although EVM is a stack-based machine, similar to JVM, Ethereum bytecode has many differences compared to the Java bytecode. For example, Java bytecode has a clearly-defined set of targets for every jump, but the jump position of Ethereum bytecode needs to be calculated during symbolic execution. Thus, \textit{Oyente} first splits opcodes into several blocks and then uses symbolic execution to build CFG (Control Flow Graph). CFG stores the relationship between blocks, e.g., jump, conditional jump. Based on the CFG, \textit{Oyente} defines several rules to detect related security issues. 

A similar method to that of \textit{Oyente} has been widely applied by other tools. For instance, \textit{GasReducer}~\citep{chen2018towards} and \textit{GasChecker}~\citep{chen2020gaschecker} are tools used to detect some gas-inefficient patterns. They use the CFG generated by \textit{Oyente}, and design patterns to detect related security vulnerability patterns. Besides, Torres et al.~\citep{torres2019art}, Chang~\citep{chang2019scompile}, Nikolic et al.~\citep{nikolic2018finding}, Zhou et al.~\citep{zhou2018security}, Krupp et al.~\citep{krupp2018teether}, Mossberg et al.~\citep{mossberg2019manticore} also use similar methods that design rules based on the CFG to detect other smart contract vulnerabilities. 

Some works make optimizations, e.g.,  Maian~\citep{nikolic2018finding} validate the results of the symbolic execution by using a concrete validation step. In the concrete validation, they create a private fork of Ethereum and then run the result generated by the symbolic execution to check its correctness. Since the results are generated by symbolic execution, and concrete validation is used to increase performance, we also classify Maian in this category. 

\noindent \textbf{Decompilers.} \textit{Vandal}~\citep{brent2018vandal} is a decompiler for smart contract bytecode. Its output includes a control-flow graph, three-address code for all operations, and function boundaries. Based on \textit{Vandal}, developers and researchers can develop other tools to maintain their smart contracts. For example,  \textit{MadMax}~\citep{grech2018madmax} uses logic-based specifications to detect gas-focused vulnerabilities of smart contracts based on the output of \textit{Vandal}.  Tsankov et al.~\citep{Securify} proposed a tool named \textit{Securify}, which uses semantic information to detect vulnerabilities of smart contracts bytecode. \textit{Securify} first decompiles the EVM bytecode. It then analyzes the data flow and control flow dependencies. Finally, it uses several patterns to check related vulnerabilities. 

\noindent \textbf{Transaction-based Tools.} \textit{TokenScope}~\citep{chen2019tokenscope} is the first tool that uses transaction histories to detect inconsistent behaviors of ERC20 Tokens. By using the stored Ethereum transaction records, \textit{TokenScope} identifies three key information of contract bytecode, i.e., core data structures, standard interfaces, and standard events. It then  compares the key information with the standard to find any inconsistent tokens. 

\noindent \textbf{Source Code Level Static Analysis.} Detecting vulnerabilities through bytecode is not easy as EVM removes some key information while compiles source code to bytecode.  \textit{SmartCheck}~\citep{tikhomirov2018smartcheck} takes smart contract source code as input, and converts the code to the AST (abstract syntax tree)~\citep{AST}. Based on the AST, \textit{SmartCheck} uses several patterns to detect 21 kinds of smart contract issues.

\subsubsection{Formal Verification}
Formal verification is a method that uses formal methods of mathematics to prove or disprove the correctness of a system~\citep{FormalVerification}. This method usually uses a formal proof on an abstract mathematical model to make the verification. 

Bhargavan et al.~\citep{bhargavan2016formal} proposed the first formal verification tool for smart contracts based on the F* proof assistant~\citep{swamy2016dependent}, and Amani et al.~\citep{amani2018towards}  presented a tool based on Isabelle/HOL~\citep{nipkow2002isabelle}. However, both of these the tools only use incomplete semantics of EVM, which might lead to errors. Thus, Park et al.~\citep{park2018formal} use a complete and thoroughly tested formal semantics of EVM to enhance the efficacy of their tool.

Kalra et al.~\citep{kalra2018zeus} introduced 11 kinds of vulnerabilities of smart contracts and proposed a tool named \textit{Zeus} to detect seven of them. \textit{Zeus} takes source code as input and translates the Solidity source code to LLVM bytecode~\citep{llvm}. Based on the LLVM bytecode, \textit{Zeus} designs several policy violations and uses a verifier to determine assertion violations.

\subsubsection{Fuzzing}
Fuzzing for smart contracts is an automated testing technique which uses random, unexpected, or invalid data as the input to the contract. Such input data is expected to lead to detecting some unwanted behaviors, e.g., crashes, failure of some functions, permission errors.  

Jiang et al.~\citep{jiang2018contractfuzzer} proposed the first fuzzing tool named \textit{ContractFuzzer}, which applies fuzzing to detect seven kinds of security issues. \textit{ContractFuzzer} utilizes smart contract ABI~\citep{Solidity} to generate fuzzing inputs. Then, they define test oracles and use static analysis to log smart contracts runtime behaviors. Finally, \textit{ContractFuzzer} analyzes the logs to find security issues. The following works make some optimization. For example, \textit{sFuzz}~\citep{nguyen2020sfuzz} can cover more branches to find more security issues. \textit{EthRacer}~\citep{kolluri2019exploiting} can run directly on Ethereum bytecode and without the need of ABI, which enlarges the usage scenario. ReGuard~\citep{liu2018reguard} provides a web service for developers to make it is easy to use. EVMFuzz~\citep{fu2019evmfuzz} designs a differential fuzz testing framework, which supports different programming languages for EVM smart contracts.

\subsubsection{Machine Learning}
With the development of the Ethereum ecosystem, some developers have used machine learning to help maintain smart contracts. Machine learning related methods need a ground truth to train the model. \textit{S-gram}~\citep{liu2018s} uses \textit{Oyente} to obtain the ground truth and utilizes a combination of N-gram language modeling and lightweight static semantic labeling to predict potential vulnerabilities. \textit{SmartEmbed}~\citep{gao2020checking} uses \textit{SmartCheck} to label the vulnerabilities and utilizes deep-learning to train the model to predict smart contract vulnerabilities. Tann et al.~\citep{tann2018towards} use \textit{MAIAN} to label the security issues and use LSTM to predict potential issues. Huang et al.~\citep{huang2018hunting} first translate the bytecode into RGB color. Based on a manually labeled dataset, they use a convolutional neural network to train the model and predict the security issues.

\subsubsection{Other Approaches}
Mavridou et al.~\citep{mavridou2018designing} proposed a tool, named \textit{FSolidM}, to automatically generate smart contracts. They claim that the generated contracts are bug-free and can reduce development efforts. \textit{FSolidM} regards smart contracts as finite state machines (FSMs). Based on FSMs, they provide a set of plugins that contain common contract design patterns and a graphical interface. Developers can add plugins to the contracts to improve security and functionalities. 

Wu et al.~\citep{li2019musc} use mutation testing to enhance the security of smart contracts. Mutation testing is a type of white-box software testing technique that changes some statements of the code and check if the test cases can find some errors. This method is based on well-defined mutation operators, and the mutation operators only make minor changes to the programs. Wu et al. designed 15 mutation operators, e.g., variable units, keywords, and use them to find bugs on smart contracts.

\subsection{Online Checking Methods}
Online checking methods can help smart contract developers defend their contracts against attacks even after they have been deployed. Table~\ref{tab:online} introduces seven publications that use online checking methods to help maintain smart contracts. However, most of the online checking methods cannot be used directly and need to be merged into the EVM if an EIP\footnote{Ethereum Improvement Proposals (EIPs) describe standards for the Ethereum platform, including core protocol specifications, client APIs, and contract standards.}~\citep{EIP} adopts any of those in a new version. 

Ayoade et al.~\citep{ayoade2019smart} proposed a method that can automatically detect vulnerable EVM bytecode segments and uses a guarded bytecode segment to replace it. Their tool is based on predefined policy rules and can only support a limited number  of simple rules. Similarly, \textit{ContractLarva}~\citep{azzopardi2018monitoring}  insert protection code into the source code of smart contracts. This updated bytecode can defend against related attacks. 

\textit{TownCrier}~\citep{zhang2016town} and \textit{FSFC}~\citep{wang2020fsfc} provide approaches to detect malicious input to protect smart contracts.  \textit{TownCrier} can be regarded as a bridge between the smart contracts and front-end programs, e.g., websites. When a frond-end program sends transactions to smart contracts, \textit{TownCrier} uses a combination of Software Guard Extensions~\citep{costan2016intel} and Intel’s recently released trusted hardware capability~\citep{intel} to check whether the input data can be trusted. \textit{FSFC} is a filter-based security framework for smart contracts. It uses several firewall rules and uses a monitor to identify malicious input. 

\textit{{\AE}GIS}~\citep{ferreira2019aegis} and \textit{VULTRON}~\citep{wang2019vultron} detect and reverse malicious transactions to protect smart contracts. \textit{{\AE}GIS} uses predefined patters to identify malicious transactions. \textit{VULTRON} compares the actual transferred Ethers and the normal transfered Ethers to find malicious transactions. 

\textit{Sereum}~\citep{rodler2018sereum} monitors state updates of smart contracts, such as changes to storage variables, to detect re-entrancy attacks. There are two components of \textit{Sereum}, i.e., a taint engine and an attack detector. 
\textit{Sereum} focuses on conditional jumps and the data that influences the conditional jumps. The taint engine is used to detect the change of state update, which loads to conditional jumps. When a re-entrancy attack happens, the state will be updated multiple times. Once the attack detector detects such malicious behaviors, the transaction will be reversed. 

\textit{ContractGuard}~\citep{wang2019contractguard}  is the first intrusion detection system for smart contracts against attacks. It monitors the network for abnormal behaviors. To detect abnormal behaviors, \textit{ContractGuard} deploys smart contracts on a testbed and trains a model. When malicious activities are detected, \textit{ContractGuard} will reverse the transactions to recover the contract states and raise an alarm to the contract owner. 

\subsection{Other Methods}
Colombo et al.~\citep{colombo2018contracts} introduced a specification-driven method that uses the \textit{DELEGATECALL} instruction to upgrade smart contracts when unwanted behaviors are detected. To detect unwanted behaviors, they predefined several checkpoints for smart contracts. The checkpoints monitor the important state of smart contracts, e.g., its balance. When an unexpected behavior is detected, the checkpoints will revert the transactions to ensure the safety of the contracts. Finally, developers are required to upgrade contracts by using the \textit{DELEGATECALL} instruction.

Marino et al.~\citep{marino2016setting} defined several standards for smart contracts and suggested developers add a \textit{Selfdestruct} function in the contracts. When the contract is attacked, the developers can undo the contracts. A similar suggestion is given by Chen et al.~\citep{contractDefects}. They suggest developers add an interrupter in the contracts. Interrupter is a mechanism to stop the contract when unwanted behaviors are detected, and \textit{Selfdestruct} function is an easy way to stop the contract.

\section{Threats To Validity}
\label{sec:threats}

\subsection{Internal Validity}

In this paper, we answered two research questions by performing a literature review. Most of the papers (74.05\%) are published between 2017 to 2019, and their findings and studies may be outdated as the Ethereum ecosystem is fast-evolving. For example, Solidity, the most popular programming language for smart contracts, has 80 versions from Jan. 2016 to Jun. 2020~\citep{Solidity_Release}. Thus, it is likely that some findings and results in the publications are out-of-date. To reduce this threat, we used an online survey to collect the opinion from many real-world smart contract developers. We compared our literature review findings with the feedback from developers to help ensure the overall validity of our findings. 

It is possible that the respondents to our survey may provide some dishonest or unprofessional answers. To reduce this influence, we first informed developers that we will not collect personal information when sending the invitation emails. The survey is anonymous and we cannot trace their information if they do not leave their email address. All questions are optional, which means developers can choose to answer a part of the questions. According to Ong et al.'s~\citep{ong2000impact} work, confidentiality and anonymity are useful to obtain un-biased data from survey respondents. 

To collect more responses, we translated our survey into a Chinese version to address the language barrier and as Google cannot be visited in China. There might be inconsistency between the Chinese and English versions of our surveys. Besides, all the respondents are written in Chinese, which needs to translate to English when analyzing the data. This process also might lead to some errors. To reduce this risk, two Chinese authors with good English skills read the survey and responded several times to ensure the correctness of the translation. 

\subsection{External Validity}
We collected responses to our survey by sending emails to Github developers. However, we might have missed some other developers who might have different opinions. Fortunately, the survey results show that the respondents to our survey have a wide variety of  backgrounds in terms of experience in developing smart contracts, job roles, and open source projects they contribute to. Thus, the diversity of backgrounds help  us to trust the survey results and can reflect real-world situations of Ethereum smart contract development. 

In the future, new functionalities will be added to Ethereum and Solidity. They might also be updated to help better address some smart contract maintenance issues. Thus, some findings and results in this paper might be out-of-date in the future. This is an inevitable trend for smart contract related empirical studies. While the methods we have identified are still working, our findings can help developers and researchers.

\section{Discussion}
\label{sec:rq3}

In this section, we discuss some future research directions and give suggestions for both developers and researchers according to our RQ1 and RQ2 findings presented in Section~\ref{sec:rq1} and \ref{sec:rq2}. 

\subsection{Improving the Smart Contract Ecosystem}


\noindent {\bf DApp Store and Comment System.}  Although there are some DApp stores for smart contracts, none of them have a smart contract verification system. They neither reject cloned contracts, nor have a rating system. As we discussed in RQ1, many copy-paste vulnerabilities are prevalent in the Ethereum blockchain's deployed smart contracts. There are also many useless smart contracts i.e. ``dead" contracts in Ethereum. These contracts are the noisy data on the blockchain and increase the difficulty of finding useful smart contracts. According to our survey, 139 (84.24\%, Q14) developers have positive opinions about the need for a DApp store like the Android Google Play Store. Such a DApp store could regulate the behaviors of smart contracts. For example, rejecting copied contracts, rating useful contracts, giving various classifications for contracts. Thus, developers could more easily find high quality contracts for reference or for use as callee contracts. A review system would allow smart contract users to submit reviews when they find bugs or suggest features that need to be improved. Such comments can help developers better maintain their contracts. It could also be a valuable research dataset. Based on such a dataset, many traditional MSR methods can be applied to help improve and maintain smart contracts.  For example, as we introduced in the previous section,  there are five machine learning-based methods to help maintain smart contracts. However, four of them use other tools to label the ground truth, and there are many false positives / negatives of the tools were used to label the ground truths. Thus, the performance of these tools is not very good. Real-world produced data, e.g., review comments,  could substantially improve the performance of these machine learning tools, just as it has for many traditional software maintenance activities and tools.

\noindent {\bf Call for High-Quality Standards, Libraries and Reference Code.} Although Ethereum has had a rapid improvement in its ecosystem, developers still claim there is a lack of standards, libraries, and useful reference code. Currently, most of the standards are published on EIPs~\citep{EIP}, and many teams provide libraries and referee code, e.g., OpenZepplelin Contracts~\citep{openzepplelin}, Smart contract best practice~\citep{bestpractice}. However, the number is still small and not enough for the vast Ethereum ecosystem.

\noindent {\bf More User Friendly Tools.} In previous sections, we introduced 41 works which can help maintain smart contracts. However, according to our survey, 96 (66.2\%, Q8) respondents claim they cannot find useful tools to debug / test / audit their contracts, or such tools are too hard to use or deploy in real-world smart contract development. An important reason for this inconsistency is that most current tools are not easy to  use for practitioners. Thus, making these tools easier to deploy and use is an important task for the future. For example, merging some tools into smart contract IDEs, or adding a user interface to the tools. 

\subsection{Improving Ethereum and Solidity}


\noindent {\bf Merging Cutting-Edge Technologies.} The previous section introduced eight online checking methods that can improve the security and maintainability of smart contracts after they have been deployed. However, most of the online checking methods cannot be used directly. Specifically, transaction detection methods can revert malicious transactions only if they were merged into the EVM and updated by nodes on Ethereum. Then, a node (minor) could revert malicious transactions instead of broadcasting to the whole Ethereum network. Similar to bytecode rewritten tools, these methods can fix a buggy bytecode snippet after they are deployed. However, this kind of method requires modification of the code stored on the blockchain, which cannot be worked directly. To use such a method, there should be a well-thought-out plan to ensure the correctness of smart contracts and the concerns of breaking the immutability (discussed in Section~\ref{sec:modification}). For example, there could be a DAO (Decentralized Autonomous Organization) responsible for updating code every a certain time by using the bytecode rewritten tools. When the DAO detects a smart contract needs to modify its bytecode, the DAO should inform the contract users / owners and allow them to vote to decide whether the code should be updated.   According to our survey (Q15), 150 (90.9\%) respondents hold positive opinions about merging cutting-edge technologies into the EVM and updated by nodes on Ethereum.

\noindent {\bf Mitigating Scalability Issues.} The scalability issue is one of the main challenges for smart contract maintenance. Several methods have been proposed to help redesign Ethereum to mitigate this issue. First,  the \textit{sharding technology} is a future direction for Ethereum to address the scalability issues. Currently, all the nodes on Ethereum need to process every transaction, which leads to low throughput. By applying sharding to Ethereum, the whole network can be split into several smaller parts, called shards. A subset of the total miner nodes would only process transactions on a certain shard. Thus, it can improve the throughput of Ethereum multiple times. Such sharding technology can also enable a smart contract to be executed by multiple threads. A contract could then be split into several parts and executed by different nodes. Enlarging the maximum stack sizes and reduce the gas cost of the storage can also mitigate the scalability issues. This mechanism aims to reduce the bulky problems of Ethereum, where all the nodes store the whole blockchain data. If the bulky problem is addressed, it is not difficult to make an optimization for stack size, database performance, and price for storage. Bruce et al.~\citep{bruce2014mini} proposed a new data structure named an account tree. The account tree holds the balance of all non-empty addresses, which enables us to remove old transactions. Thus, new nodes do not need to store all transactions and can reduce the total bulk of the blockchain. 


\noindent {\bf Trusted Modification Methods.}  In Section~\ref{sec:modification}, we introduced four modification methods for smart contracts. Among them, using the \textit{Selfdestruct} function and developing \textit{upgradeable contracts} cost the least. However, these two methods can lead to a major trust concern from the users and other security issues. Previous work~\citep{Chen-selfdestruct} introduced a method to reduce the trust and security concern for the usage of the \textit{Selfdestruct} function, which can also be applied to upgradeable contracts. This method suggests that developers should distribute the rights to the users of the contracts. They could vote to decide whether the contracts should be destructed or upgraded. Using consensus protocols, such as PoS~\citep{PoS}, DPoS~\citep{DPoS} are examples of such voting. For example, if a user invests 100 Ethers to the contract, the user has 100 score to vote. The more Ethers users invest contracts, the more rights they have. When the voting process finished, users who do not agree can transfer their Ethers to other accounts. Also, the delay can reduce the risk of the Ethers locking, as Ethers transferred to the destructed address will be locked forever. During the voting and delaying steps, developers should suspend the function of the contracts to prevent attacks or other unwanted behaviors. 

\section{Related Work}
\label{sec:related}

We review previous key empirical studies on smart contracts, and highlight the difference between our work at the end of the section. 
\subsection{Survey Based Smart Contract Empirical Studies}

Bosu et al.~\citep{bosu2019understanding} pre-designed some questions and used an online survey to collect the opinions from developers on Github. Their work aimed to answer who contributes to smart contracts and their motivation for development, what is the difference between smart contract development and traditional software development, the challenges of smart contract development, and what kinds of tools that developers feel they need. 


Chakraborty et al.~\citep{chakraborty2018understanding} sent an online survey to 1,604 developers on Github and received 145 responses. Their survey aimed to find the best current software development practices for smart contracts. Their findings suggest that some traditional software engineering practices are still working for blockchain projects. They identified that the smart contract ecosystem is immature and needs more SE methods, resources, and tools.


Chen et al.~\citep{contractDefects} defined 20 contract defects by analyzing posts on Stack Exchange. They divided the defects into five categories, i.e., security, availability, performance, maintainability, and reusability defects. They claimed that removing these contract defects can improve the robustness and enhance development efficiency. To validate whether real-world developers regard these contracts as harmful, they use an online survey to collect developers' opinions. The results show that all the 20 contract defects are potential harmful to smart contracts.

\noindent {\bf Our Novelty and Difference: } Both our work and Bosu et al.'s work~\citep{bosu2019understanding} investigated the challenges of smart contracts development. Our work investigated the maintenance-related challenges for post-deployed Ethereum Smart Contract development, which is much more comprehensive than Bosu et al.'s work. The only similarity between two works is that we both reported the lack of tools is one of the challenges for smart contract development / maintenance, while our work has a deeper analysis for the reasons why the academia proposed many tools with excellent performance but the smart contract developers also feel they lack tools to check smart contract security.  (See Section 4.2.1). 

There is a big difference between Chakraborty et al.'s work~\citep{chakraborty2018understanding} and our work. Their work aims to understand the software development practices of smart contract projects. For example, how smart contract developers test their code; what's the requirement during the development, while our work focuses on the challenges during smart contract development. Both of our works used surveys to collect developers' opinions; their work used surveys to find the answers of pre-defined research questions, while our survey aimed to validate the findings that we collected from literature reviews.  

Chen et al.'s work~\citep{contractDefects} reported detailed patterns / code that are harmful for smart contract development / maintenance, while our work stood at a higher level that reports the challenges of smart contract maintenance instead of detailed code patterns.   

\subsection{ Literature Review Based Smart Contract Empirical Studies.} 
 Conoscenti et al.~\citep{conoscenti2016blockchain} proposed an empirical study to help developers understand how to use smart contracts and blockchain technology to build a decentralized and private-by-design IoT system. To obtain key related information they conducted a systematic literature review based on 18 publications. Their work introduced several use cases of blockchain in the IoT domain and the factors affect integrity, anonymity, and adaptability of blockchain technology.

Udokwu et al.~\citep{udokwu2018state} selected 48 publications from 496 papers. Based on the selected papers, they described the key current usages of smart contract technology and challenges in adopting smart contracts to other applications. Their analysis showed that the most popular applications of smart contracts are supply chain management, finance, healthcare, information security, smart city, and IoT. They also identified 18 limitations of blockchain technology that affects the adoption of smart contracts for other applications.

Macrinici et al.~\citep{macrinici2018smart} pre-defined seven research questions and selected 64 publications to find  answers. Their results show that the most popular topic in smart contract research is offering solutions to address related problems, e.g, developing tools, proof-of-concepts, and designing protocols. They also summarized 16 smart contract related problems and divided them into three categories, i.e., blockchain mechanism, contract source code, and EVM problems. 

\noindent {\bf Our Novelty and Difference: } Our work is the most comprehensive literature review based on smart contract empirical study (our 131 publications v.s.  Conoscenti et al. 's 18 publications v.s. Udokwu et al. 's 48 publications v.s. Macrinici et al. 's 64 publications). Besides, this is the only work that uses an online survey to validate our findings from the literature review. There might be a gap between academia and industry knowledge, usage, practices, and desired outcomes. Also, the fast-growing ecosystem of Ethereum can also create errors for some even recent findings.  Thus, the findings based exclusively on literature reviews might not be reliable. For example, Zhou et al.~\citep{zou2019smart} mention that Solidity lacks the support of \textit{try-catch}, which increases the difficulty of the development. However, Solidity adds this support from version 0.6.0~\citep{Solidity}. Also, this work is the only one that focuses on smart contract maintenance issues, while the mentioned three works focus on IoT, adopting smart contracts to other applications, and the most popular topic in smart contract research, respectively. 

\subsection{ Security Related Smart Contract Empirical Studies.} 
Li et al.~\citep{li2017survey} reviewed security issues for the blockchain systems from 2015 to 2017. They classified these issues into nine categories and introduced the related causes. For example, one of the categories is the ``51\% vulnerability" and the cause is the consensus mechanism. To help developers understand such attacks better, they also gave example real attacks as case studies and analyzed the vulnerabilities utilized by the attackers. 

Bartoletti~\citep{bartoletti2020dissecting} found that the infamous Ponzi scheme has migrated to Ethereum. Misbehaving developers use smart contracts to design a Ponzi scheme to make money. Bartoletti et al. manually checked real-world smart contracts and summarized four kinds of Ponzi smart contracts, i.e., tree-shaped, chain-shaped, waterfall, handover Ponzi scheme. To help further research on Ponzi scheme detection, they manually labeled a dataset that contains 184 schemes.  A follow-up work~\citep{ponzi-www, chen2019exploiting} used this dataset to design machine learning methods to detect Ponzi smart contracts. 

Delmolino et al.~\citep{delmolino2016step} are the lectures of a university who teach smart contract programming. They documented the pitfalls of smart contracts according to their teaching experiences. The pitfalls include errors in encoding state machines, failing to use cryptography, misaligned incentives, and Ethereum-specific mistakes. 

Atzei et al.~\citep{atzei2017survey} studied attacks on smart contracts on Ethereum between 2015 to 2017, and provided a classification of programming pitfalls which might lead to the security issues of smart contracts. Their work introduced six vulnerabilities in the Solidity level, three vulnerabilities in the EVM level, and three vulnerabilities in the blockchain level. For most of the vulnerabilities introduced in the paper, a detailed introduction, code examples, and attack examples are given to help readers better understand. 

\noindent {\bf Our Novelty and Difference: } The motivation between our work and these security-related smart contract empirical studies have big differences. Our work aims to highlight the maintenance-related concerns for post-deployed Ethereum smart contract development, and security concerns is only a very small part of our work. These works focus on exactly security issues with more detailed information, e.g., the detailed code patterns and attack examples. 

\subsection{ Other  Smart Contract Empirical Studies.}  Zheng et al.~\citep{zheng2020overview} described the challenges of developing smart contracts in the whole life cycle, including creation challenges, deployment challenges, execution challenges, and completion challenges. Their work not only focused on the Ethereum platform, but is also more narrow in other ways. Thus, they also analysed some differences between six smart contract platforms. Another work~\citep{zheng2018blockchain} discussed the challenges of the blockchain system, and the opportunities of blockchain technology. For the challenge, they mainly focused on the architecture of blockchain and consensus algorithms. For the opportunities, they introduced the applications of blockchain, e.g., IoT, Finance. Reyna et al. ~\citep{reyna2018blockchain} investigated the challenges of applying blockchain technology to the IoT to increase the security and reliability. Mohanta~\citep{mohanta2018overview} introduced seven uses cases for smart contracts, including supply chain, IoT, and healthcare systems. Many empirical studies also focus on the performance of smart contract tools~\citep{perez2019smart, parizi2018empirical}, programming languages~\citep{harz2018towards, schrans2018writing, parizi2018smart}, ecosystem~\citep{kiffer2018analyzing, he2019characterizing, hegedHus2019towards}, permissions~\citep{vukolic2017rethinking}, design patterns~\citep{bartoletti2017empirical}, life cycle~\citep{di2019mayflies}, call relations~\citep{bistarelli2019analysis}. Durieux et al.~\citep{durieux2020empirical} presented an empirical study of 9 state-of-art smart contract vulnerability analysis tools. To evaluate the tools, they use two datasets, i.e., a small-scale dataset consists of 69 vulnerable smart contracts and a large-scale dataset with all verified smart contracts (47, 518 contracts) on Etherscan.  They found that only 42\% of vulnerable smart contracts in small-scale dataset can be detected by all the 9 tools. About 97\% of smart contracts are labeled as vulnerable by at least one tool. According to their analysis result, Mythril~\citep{Mythril} has the highest accuracy (27\%) in detecting smart contract vulnerabilities. 

\noindent {\bf Our Novelty and Difference: } 
In this paper, we summarized the key \emph{maintenance} issues and current maintenance methods for smart contracts as evidence from our literature review, which has a different topic with the smart contract empirical studies mentioned above. This is also the only work to date that has conducted a literature review to collect maintenance issues of smart contracts and used an online survey to validate these findings with practitioners.

\section{Conclusion}
\label{sec:conclusion}

In this paper, we conducted the first empirical study on the Ethereum smart contract maintenance issues. We performed a systematic literature review to obtain related information and used an online survey to validate our findings with practitioners. Our study contains two research questions.  In RQ1, we identified 9 kinds of issues related to corrective, adaptive, perfective, and preventive maintenance of smart contacts, and another 4 issues corresponding to the overall maintenance process for smart contracts. In RQ2, we summarized current maintenance methods used for smart contracts from 41 publications and divided them into three categories, offline checking methods, online checking methods, and other methods. We also highlighted two kinds of future research directions and discussed some suggestions for both smart contract developers and researchers according to the previous RQ answers and our survey results. 


\section*{Acknowledgements}

Grundy is supported by ARC Laureate Fellowship FL190100035. This work is partially supported by ARC Discovery Project DP200100020.

\balance
\bibliographystyle{spbasic}
\bibliography{ref}

\end{document}